\documentclass{aa}
\usepackage{natbib}
\usepackage{graphicx}
\usepackage{appendix}
\usepackage{txfonts}
\usepackage{dcolumn}
\usepackage{arydshln}
\usepackage{color}

\begin{document}

   \title{New constraints on the structure of the nuclear stellar cluster of the Milky Way from star counts and MIR imaging}


   \author{E. Gallego-Cano
          \inst{1,2}
          \and
          R. Sch\"odel
          \inst{2}
          \and
          F. Nogueras-Lara
          \inst{2}
          \and
          H. Dong
          \inst{2}
          \and
          B. Shahzamanian
          \inst{2}
          \and
          T. K. Fritz
          \inst{3,4}
          \and
          A. T. Gallego-Calvente
          \inst{2}
          \and
          N. Neumayer
          \inst{5}
          }

   \institute{
    Centro Astron\'omico Hispano-Alem\'an (CSIC-MPG), Observatorio Astron\'omico de Calar Alto, Sierra de los Filabres, 04550, G\'ergal, Almería, Spain
             \email{egallego@caha.es}
        \and
    Instituto de Astrof\'isica de Andaluc\'ia (CSIC),
     Glorieta de la Astronom\'ia s/n, 18008 Granada, Spain
         \and
      Instituto de Astrof\'isica de Canarias, Calle V\'ia L\'actea s/n, E-38205 La Laguna, Tenerife, Spain
         \and
      Universidad de La Laguna, Dpto. Astrof\'isica, E-38206 La Laguna, Tenerife, Spain
        \and
      Max-Planck Institute for Astronomy, K\"onigstuhl 17,69117 Heidelberg, Germany
                 }

\abstract
{The Milky Way nuclear star cluster (MWNSC) is a crucial laboratory for studying the galactic nuclei of other galaxies, but its properties have not been determined unambiguously until now.}
 { We aim to study the size and spatial structure of the MWNSC.}
 { This study uses data and methods that address potential shortcomings of previous studies on the topic. We use $0.2''$ angular resolution $K_{s}$ data to create a stellar density map in the central $86.4$pc x $21$pc at the Galactic center. We include data from selected adaptive-optics-assisted images obtained for the inner parsecs. In addition, we use IRAC/Spitzer mid-infrared (MIR) images. We model the Galactic bulge and the nuclear stellar disk in order to subtract them from the MWNSC. Finally, we fit a S\'ersic model to the MWNSC and investigate its symmetry.}
{Our results are consistent with previous work. The MWNSC is flattened with an
axis ratio of $q=0.71 \pm0.10$, an effective radius of $R_{e}=(5.1
\pm1.0)$ pc, and a S\'ersic index of $n=2.2 \pm0.7$. Its major axis may be
tilted out of the Galactic plane by up to $-10$ degrees. The distribution of the
giants brighter than the Red Clump (RC) is found to be significantly flatter than the distribution of the faint stars. We investigate the 3D structure of the central stellar cusp using our results on the MWNSC structure on large scales to constrain the deprojection of the measured stellar surface number density, obtaining a value of the 3D inner power law of $\gamma=1.38\pm0.06_{sys}\pm0.01_{stat}$.}
{ The MWNSC shares its main properties with other extragalactic NSCs found in spiral galaxies. The differences in the structure between bright giants and RC stars might be related to the existence of not completely mixed populations of different ages. This may hint at recent growth of the MWNSC through star formation or cluster accretion.}


  \titlerunning{New constraints on the structure of the MWNSC}
   \maketitle
%

\section{Introduction}

Nuclear stellar clusters (NSCs) are the densest and most massive star clusters in the universe, with half-light radii of $\sim 2-8$ pc and masses of $\sim10^{6}-10^{7}$M$_{\odot}$ \citep{Boker:2002kx,Boker:2004oq,Walcher:2005ys,Cote:2006eu,Georgiev:2014ve}. They follow similar scaling relations to their host galaxies in a similar way to super massive black holes
(SMBHs), suggesting a common formation mechanism of NSCs
and SMBHs closely linked to galaxy evolution. The
formation mechanisms of NSCs are still debated.
There are two formation scenarios: infall and subsequent
merging of star clusters
\citep[e.g.,][]{tremaine75, Antonini:2013ys, Antonini:2014aa, Arca-Sedda:2014aa} and in situ formation of stars at the center of a
galaxy \citep[e.g.,][]{Milosavljevic:2001ly, McLaughlin:2006fk, Brown:2018aa}. To bring some clarity to this topic we need to study the
structure, stellar populations, and kinematics of
NSCs. The main limitation for detailed studies of NSCs is
their compactness, which makes
characterisation of their resolved structure challenging for galaxies at large distances.

However, the center of the Milky Way is located at only $\sim8$ kpc from Earth \citep{Ghez:2008fk,Gillessen:2009qe,2018A&A...615L..15G}. It has a $4\times10^{6}$\,M$_{\odot}$
\citep[e.g.,][]{Boehle:2016zr,2017ApJ...837...30G,2018A&A...615L..15G} massive black hole,
Sagittarius\,A* (Sgr\,A*), surrounded by a $\sim$ $2.5\times10^{7}$\,M$_{\odot}$
nuclear star cluster
\citep{Schodel:2014fk,Schodel:2014bn,Feldmeier:2014kx,Chatzopoulos:2015yu,
Feldmeier-Krause:2017rt} with a half-light radius or effective radius of
$\sim 4.2-7$ pc
\citep{Schodel:2014fk,Schodel:2014bn,2016ApJ...821...44F}. The MWNSC was
discovered by \cite{Becklin:1968nx} as a dense stellar structure with a
diameter of a few arcminutes and a mass of a few $10^{7}$\,M$_{\odot}$.
It lies embedded within the nuclear stellar disk (NSD), a flat disk-like
structure with a radius of $230$ pc and scale height of $\sim45$ pc
\citep[][]{Launhardt:2002nx}. The MWNSC is the only galactic nucleus
where we can resolve the stars observationally on scales of about
2\,milliparsecs (mpc), assuming diffraction-limited observations at
about $2\,\mu$m with an 8-10m telescope. It is therefore a unique test
system to study NSCs and explore the transition regimen between SMBH-dominated or NSC-dominated galaxy cores
\citep{Graham:2009lh,Neumayer:2012fk}.

Although the MWNSC is so close to us, studies of this cluster must overcome various difficulties. Surrounded by miscellaneous components, such as the Galactic bulge (GB), the NSD, and the Galactic disk (GD), it is a challenge to {isolate}. The extreme extinction ($A_{v}\geq 30$, $A_{k}\sim 3$) toward the GC \citep[e.g.,][]{Nishiyama:2009oj,Schodel:2010fk,Fritz:2011fk} means that it can only be observed at infrared wavelengths, and the strong source crowding requires us to use the highest angular resolution possible. Moreover, the NSC presents a complex formation history \citep[see][and references therein]{Pfuhl:2011uq}. The observed ages of the stars range from $\sim12$~Gyr to a few million years. In addition, due to observational difficulties, the typical spectroscopic completeness reaches only around $K_{S}=15.5$ stars, currently  limiting any direct spectroscopic measurements to at most a few hundred stars \citep[]{Do:2009tg,Bartko:2010fk,Do:2015ve,Stostad:2015qf,Feldmeier-Krause:2015}.

 The purpose of the present paper is to
 systematically study the size and spatial structure of the NSC
 of our Galaxy, building on the works of
 \citet{Schodel:2014fk} and
 \citet{Fritz:2014vn,2016ApJ...821...44F} but using new and
 different data, in particular high-angular-resolution near-infrared (NIR) imaging on scales of several tens of parsecs.
 \citet{Schodel:2014fk} used data in the central $200$
 pc from Spitzer/IRAC data at $3.6 \mu$m and $4.5 \mu$m
  with an angular resolution of $ 2''$ that are
 affected by strong diffuse emission from the mini-spiral
 and the presence of a few extremely bright sources (e.g.,
 IRS1W or IRS7). The other comparable study is by
 \cite{2016ApJ...821...44F}, who used three different data
 sets: data from the adaptive-optics-assisted NIR camera NACO/CONICA (NACO) at the ESO Very Large Telescope (VLT) in the central $20''$ ($0.8$ pc) with
 an angular resolution of $\sim0.08''$; data from the Hubble Space
 Telescope (HST) Wide Field Camera 3 (WFC3)/IR in the central $68''$($2.72$
 pc), with angular resolution of $0.15''$; and VISTA Variables in the Via Lactea data obtained with VISTA InfraRed CAMera (VIRCAM) in the central $2000''$($80$ pc), with an angular resolution of $1.0''$.

 We use a field of view (FOV) of $86$ pc x $20.2$ pc and we use two different data sets: high resolution data with a point spread function (PSF) of $0.2''$ full width at half maximum (FWHM) from the GALACTICNUCLEUS survey \citep{2018A&A...610A..83N, 2019A&A...631A..20N} and even higher resolution data ($0.05''$) from \citet{2018A&A...609A..26G} for the inner parsecs, the most crowded region near Sgr\,A*. Moreover, we use a larger field from \citet{Nishiyama:2013uq} of the central $\sim860$ pc x $280$ pc of our Galaxy to explicitly take into account the influence of the different Galactic components that appear superposed on the line-of-sight toward the NSC. One novel aspect of this work is that we use a GB model to minimize the bias of the GB on the measured properties of the NSD and NSC. We also repeat the work of \citet{Schodel:2014fk} to obtain independent constraints on the structure of the MWNSC and even improve it by using a larger FOV and including the GB model.

 We study the structure of the MWNSC in different stellar brightness ranges to test whether the composition of the stellar population changes with distance from the center of the MWNSC and whether stars of different brightness outline the same structure. We explore in detail the systematic uncertainties that can affect the parameters, taking into account different potential sources of systematic errors. Another important aspect of our work is that we study the symmetry of the NSC assuming that it is aligned with respect the Galactic plane (GP), as previous work did, but we go one step further in exploring it by leaving the tilt angle between the NSC and the GP free in the fits. Finally, we return to the study of the stellar cusp at the Galactic center (GC) originally carried out by \citet{2018A&A...609A..26G} and \citet{2018A&A...609A..27S} by using GALACTICNUCLEUS data for the projected radii $R\gtrsim2$\,pc to recompute the 3D density structure of the stars near the massive black hole, Sgr\,A*.

 The paper is organized as follows: In section\,\ref{sec:data}, we present our data, and in section\,\ref{sec:Methodology} we describe the methodology to create the stellar number density maps. In section\,\ref{sec:structure_nsc} we model the GB and NSD, and we fit S\'ersic functions to the stellar density maps of the MWNSC. We also present our improved work on MIR images to study the MWNSC and the NSD in section\,\ref{sec:structure_nsc}. We discuss our results, and study the implications for the stellar cusp around Sgr\,A* in section\,\ref{sec:discussion}. We conclude in section\,\ref{sec:conclusions}. Finally, we explore potential sources of systematic error on the NSC best-fit parameters in Appendix\,\ref{sec:syst_2d}, and on the inner power-law index $\gamma$ that describes the cusp in Appendix\,\ref{sec:syst_nuker}.

\section{Data set\label{sec:data}}
\begin{figure*}[!htb]
\includegraphics[width=\textwidth]{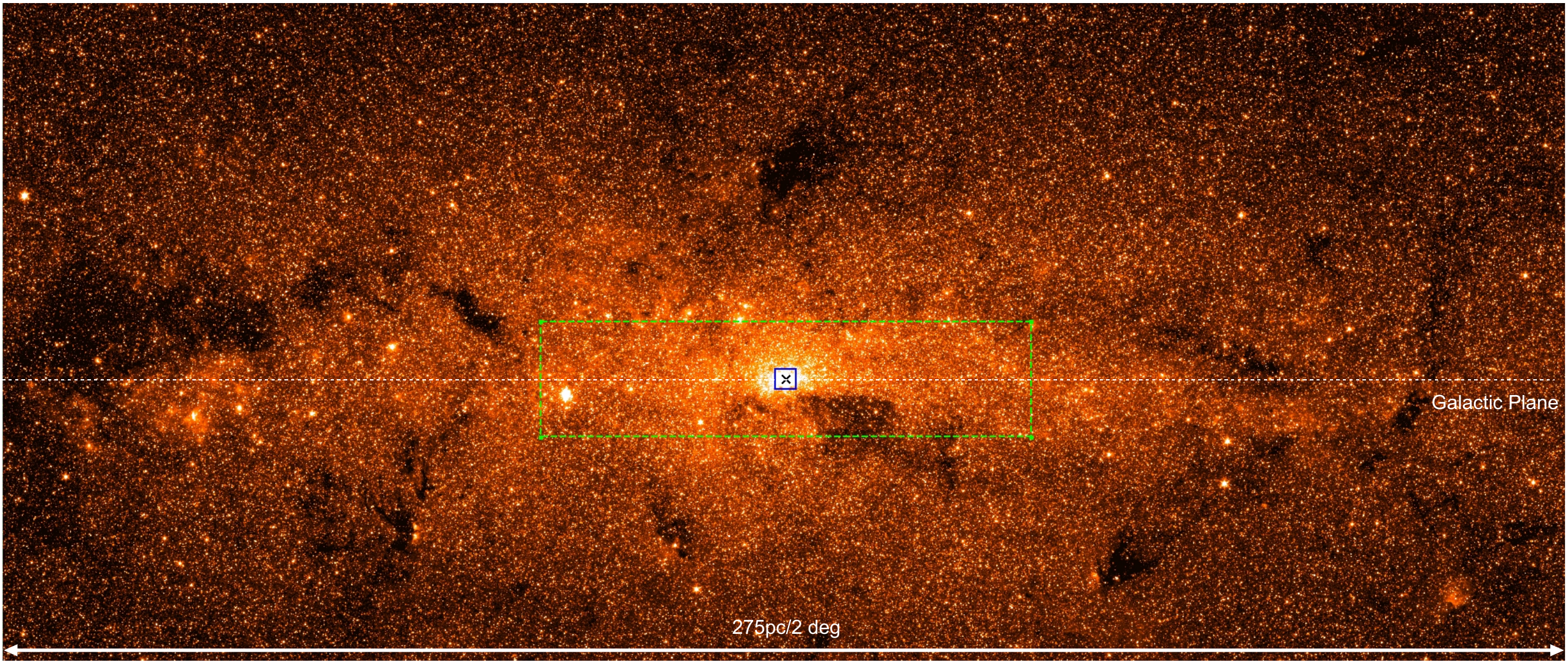}
\caption{\label{Fig:spitzer} Extinction-corrected $4.5\mu$m Spitzer/
IRAC image of the GC from \citet{Schodel:2014fk}. Galactic north is
up and Galactic east is to the left. The blue square outlines the
NACO FOV of $\sim$ $1.5'\times1.5'$ ($\sim$ $3.6$ pc $\times$ $3.6$ pc)
and the green rectangle the HAWK-I FOV of $\sim$ $35.8'$ $\times$ $8.4'$
($\sim$ $86$ pc $\times$ $20.2$ pc). Sgr\,A* is located at the position
of the black cross.}
\end{figure*}

To study the overall large-scale structure of the MWNSC we create stellar density maps using two data sets depending on the distance to Sgr\,A*. We use HAWK-I data in the central $84.4$ pc x $21$ pc of the Galaxy (see the green rectangle in Fig. \,\ref{Fig:spitzer}). Nevertheless, HAWK-I images are not complete due to the crowding in the magnitude ranges that we want to study in the central $\sim1.5$ pc of the Galaxy. For this reason, we need higher angular resolution data in this region. We use NACO data in the central two parsecs around the SMBH (see the blue square in Fig. \,\ref{Fig:spitzer}).

\subsection{HAWK-I data\label{sec:HAWKI}}
We use $Ks$-band data obtained with the near-infrared camera High Acuity Wide field K-band Imager (HAWK-I) at the ESO VLT from the GALACTICNUCLEUS survey, a JHKs imaging survey of the center of the Milky Way \citep[see ][]{2018A&A...610A..83N, 2019A&A...631A..20N}. We use the central $14$ fields of the survey described in detail in \cite{2019A&A...631A..20N} that together compose a region of about $84.4$ pc x $21$ pc (see the green rectangle in Fig. \,\ref{Fig:spitzer}) around Sgr\,A*. We do not have field 7 (black rectangle to the northeast in Fig. \,\ref{Fig:densitymaps}) because the data were obtained under poor observing conditions. Since this field is at a radial distance of more than $11$ pc from Sgr\,A*, we assume that it is not very significant for studying the NSC (half-light radius $\sim3-5$ pc).
\subsection{NACO data\label{sec:NACO}}
The data and their reduction are described in detail in \citet{2018A&A...609A..26G}; they provide an angular resolution of about $0.070''$ FWHM over a field of about $1.5'x1.5'$. In particular, we use $Ks$-band data from 11 May 2011 obtained with the S27 camera ($0.027''$ pixel scale) of NACO installed at the ESO VLT. It is the largest field observed by NACO and   is therefore the best option for studying the large-scale structure of the NSC. Even though the data are relatively shallow, they are of excellent quality and are well-suited to studying the central parsecs at the GC in the magnitude ranges of stars that we want to analyze ($K_{s}\leq16$).

\section{Stellar number density maps \label{sec:Methodology}}
The steps to create the stellar density maps are the following:
\begin{enumerate}
  \item Alignment of NACO and HAWK-I data via a two-degree polynomial fit
   (IDL POLYWARP and POLY\_2D procedures). We perform iterative fits using lists of detected stars in the images to compute the parameters.
  \item Application of an inner mask to HAWK-I data with the shape of the NACO data FOV in the central $\sim1.5$ pc.
  \item The joining of NACO and HAWK-I data by registering the astrometry of the detected sources in the overlapping region and using NACO in the inner region.
  \item Exclusion of all spectroscopically identified early-type stars, that is,\ massive and
  young ones, from our sample \citep[using the data of][]{Do:2013fk}.
  \item Application of an extinction correction to compute the intrinsic magnitudes for the stars (see section\,\ref{sec:extincion}).
  \item Selection of the range of $K_{s}$-magnitude of the stars (see section\,\ref{sec:selection}).
  \item Creation of the maps using 2D density functions (IDL HIST\_2D procedure). We assume a bin size of $5''\times5''$. (See Appendix\,\ref{sec:binning} where we study the selection of the bin size).
  \item Computation of the mean density in the overlap region between HAWK-I and NACO data.
\end{enumerate}
\begin{figure*}[!htb]
\includegraphics[width=\textwidth]{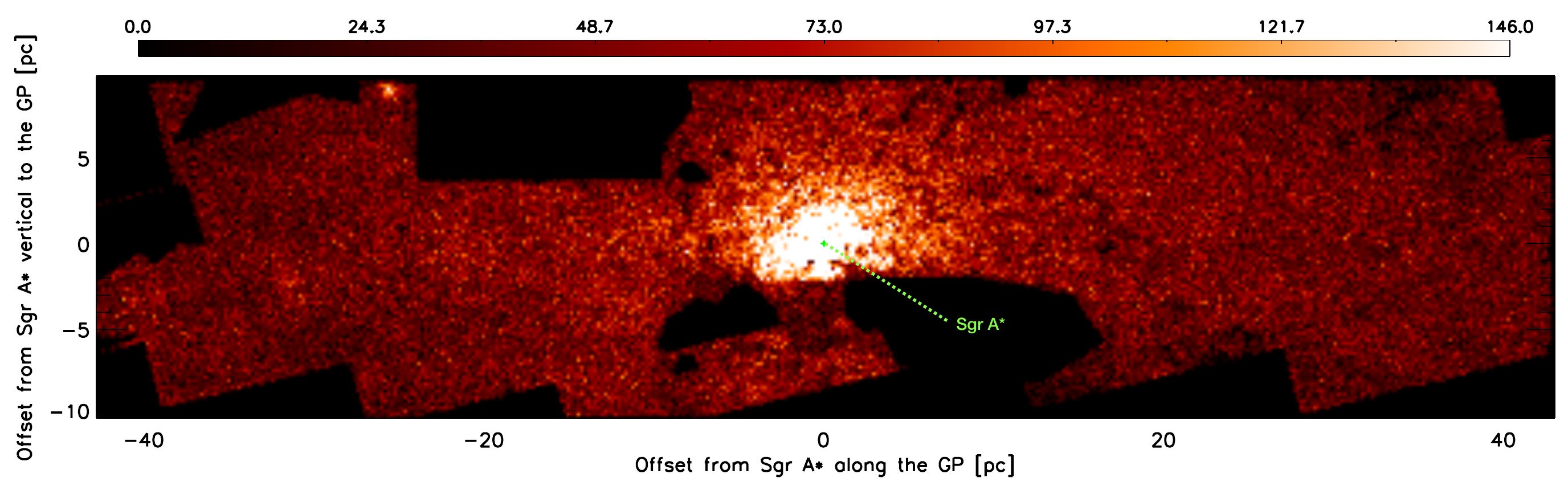}
\caption{\label{Fig:densitymaps} Extinction-corrected
stellar number density map of the central $86$ pc $\times$ $20.2$
pc of the Galaxy for stars with $11.0 \leq K_{s}\leq16.0$
($9.0 \leq K_{s, extc}\leq14.0$). The magnitudes of
individual stars were corrected for extinction using the
extinction map of \citet{Dong:2011ff}. The dark
patches in the image correspond to the green polygons in the extinction map (see Fig.\,\ref{Fig:ext_map}).
Sgr\,A* is located at the green star. Galactic north is up
and Galactic east is to the left. The Galactic plane runs
horizontally.} 
\end{figure*}
The image in Fig. \,\ref{Fig:densitymaps} shows the extinction-corrected density map for stars in the range $9.0 \leq K_{s, extc}\leq14.0$, where $K_{s, extc}$ is the extinction-corrected magnitude. 

\subsection{Crowding and completeness \label{sec:crowding}}

For HAWK-I data, we study $14$ fields that have different
quality mostly because of the different atmospheric seeing
conditions during the observations. We take special care in
analyzing the central $10$ parsecs because this is the most
crowded region in the whole field and may therefore have a
higher incompleteness. Firstly, we create masks to take
into account the inhomogeneity of the field, as well as the
different crowding in the regions. Figure\,\ref{Fig:densitymaps} shows the extinction-corrected density map for stars in the magnitude range of $9.0 \leq K_{s, extc}\leq14.0$ , where we can see regions with
different stellar number density. We use the density map in order to create two masks: a
global mask, where regions with extremely low (large IR
dark clouds to the SW) and high density (H2 regions to the
NE, and the Arches and Quintuplet clusters at
projected distances of 26–32 pc from the SMBH, respectively) are masked; and a local mask to take
into account the different completeness of the fields. For
the former, we test four different density thresholds
to mask out regions with low density due to high extinction (e.g., the two filaments to the NW). We create four global masks where 30, 40, 50, and 60\% of the pixels with low density are excluded (Mask 1 - least restrictive, to Mask 4 - most restrictive). For the latter, we
mask positions in the density maps where the values are too
low or high by comparing with their neighbors. For each
position in the density map, the mean density is computed assuming the values of the density in the closest
positions inside a ring with a radius of $15''$ ($0.6$ pc),
rejecting the points whose values are larger than 1$\sigma$
from the mean. The final masks are obtained by multiplying
both local and global masks (see more details in
Appendix\,\ref{sec:masks}).

\begin{figure}[!htb]
\includegraphics[width=\columnwidth]{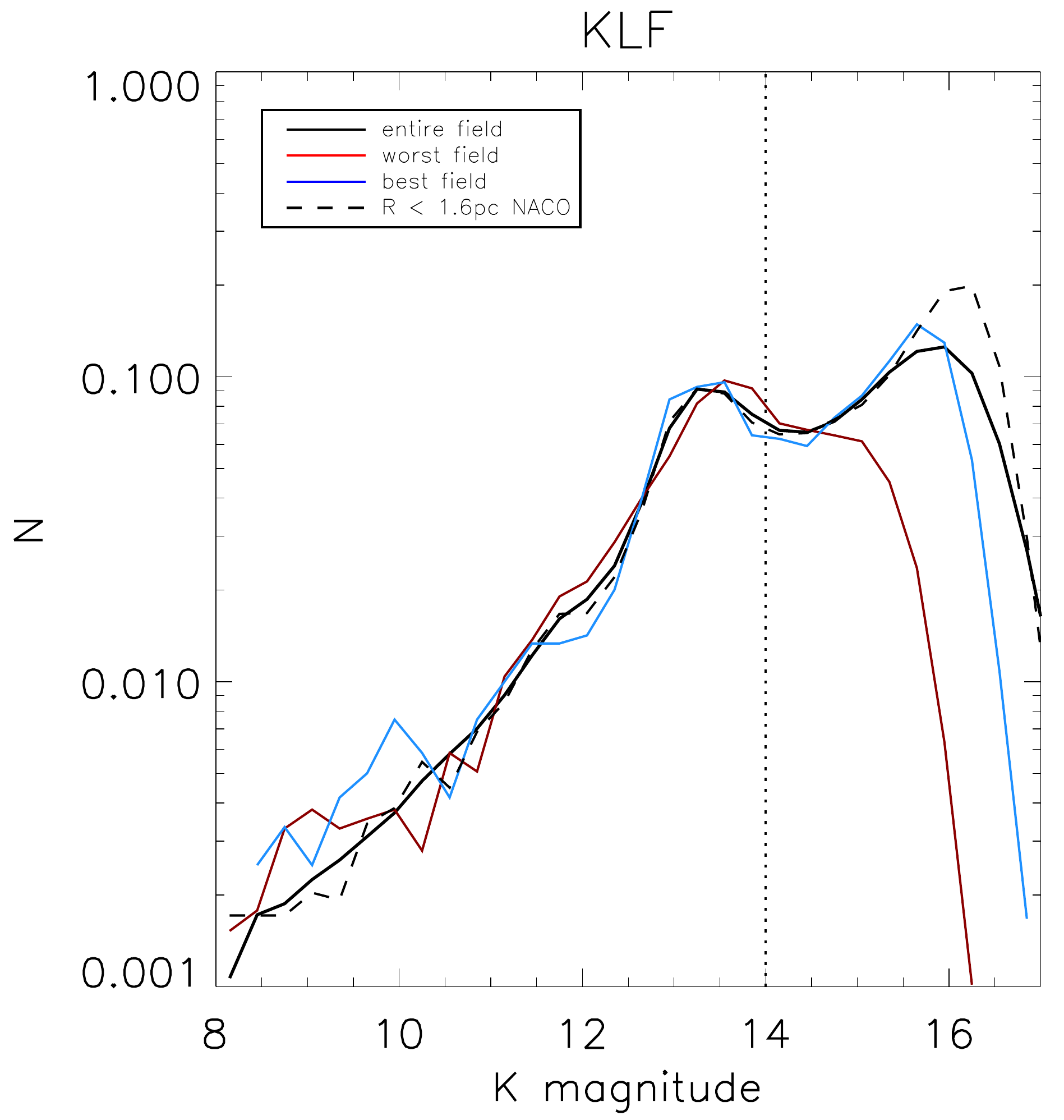}
\caption{\label{Fig:KLF_figure} Comparison between the KLFs determined from the entire field (black line) and those from the worst field (red line), the best field ( blue line), and NACO data (dashed line). The KLFs are computed after correcting the magnitude of each star using the extinction map of \citet{Dong:2011ff} and applying Mask 3. The dotted straight line indicates the faintest magnitude down to which the counts can be assumed complete for all fields.}
\end{figure}

In order to determine the faintest magnitude down to which the
counts can be assumed to be complete for all fields, we study the $K_ {s}$-luminosity functions (KLFs)
of the individual fields after applying Mask 3. Figure \,\ref{Fig:KLF_figure} shows a
comparison between the KLFs determined from our mosaic (the black line) and those from the worst field
(the red line), the best field (the blue line), and NACO
data (the dashed line). The regions where the KLFs are computed are
circular with a radius of $1.5$ pc centered at a distance of $\sim6$
pc from Sgr\,A* to the west (the worst field), and a distance of
$\sim29$ pc to the NE (the best field), respectively. The KLF from
NACO data is computed in the central $\sim1.5$ pc around Sgr\,A*. The faintest magnitude down to which the counts can be assumed
complete is $K_{s, extc}=14.0$ ($K_{s}\sim16.0$) for all fields
(the dotted straight line in Fig.\,\ref{Fig:KLF_figure}).

We do not apply a completeness correction for NACO data because the high-angular-resolution data used here are complete in the range of magnitudes that we study \citep[$K_{s}\leq16.0$, see][]{2018A&A...609A..26G}. We do not apply a completeness correction for HAWK-I data either but we apply masking generously to those regions with very low density, as we described above.

\begin{figure*}[!htb]
\includegraphics[width=\textwidth]{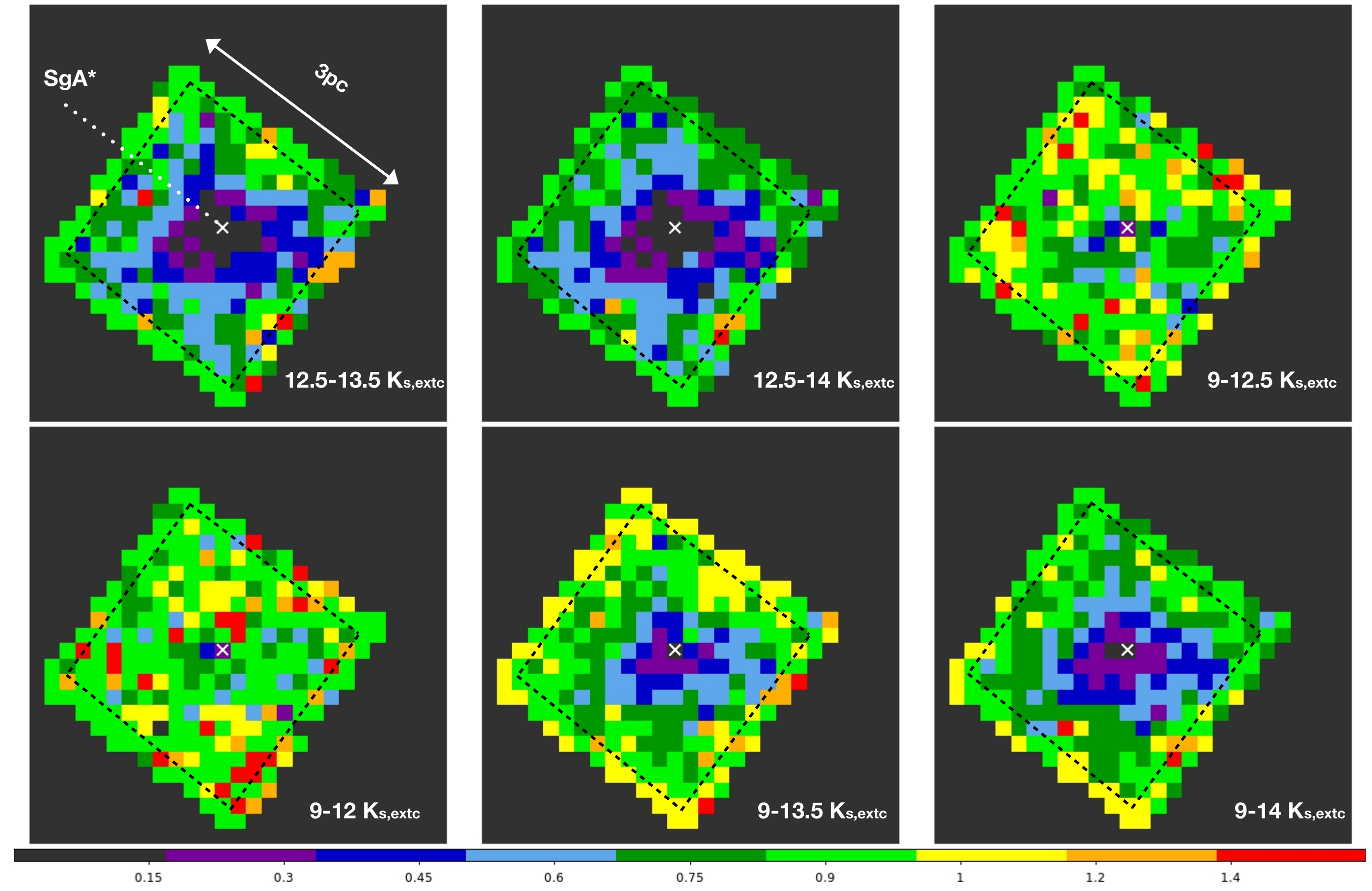}
\caption{\label{Fig:completeness} Completeness maps for HAWK-I in the central $\sim1.5$ pc of the Galaxy by comparing with NACO data. The maps represent the number of stars from HAWK-I data divided by the number of stars from NACO data. The region where only NACO data are used to create the density maps is outlined in the panels by the dashed black square. The outer region is the overlapping region, where we take the mean between both data sets. We can see that the completeness is greater than 88\% in the edges in all cases, and therefore we do not apply any scaling factor when we join the data.}
\end{figure*}

We analyze the completeness of HAWK-I in the central $\sim$1.5 pc by comparing with NACO data. We define the completeness of HAWK-I in the central region as the number of stars from HAWK-I data divided by the number of stars from NACO data. Figure \,\ref{Fig:completeness} shows the completeness maps computed for different magnitude ranges. We can see that in the overlapping region the completeness is greater than 88\% for stars brighter than $K_{s, extc}=14.0$. Therefore, we do not apply a factor of overlapping to join the data but take the mean between NACO and HAWK-I data in the overlapping region to ensure a smoothed transition between both data sets.

\subsection{Extinction \label{sec:extincion}}
\begin{figure*}[!htb]
\includegraphics[width=\textwidth]{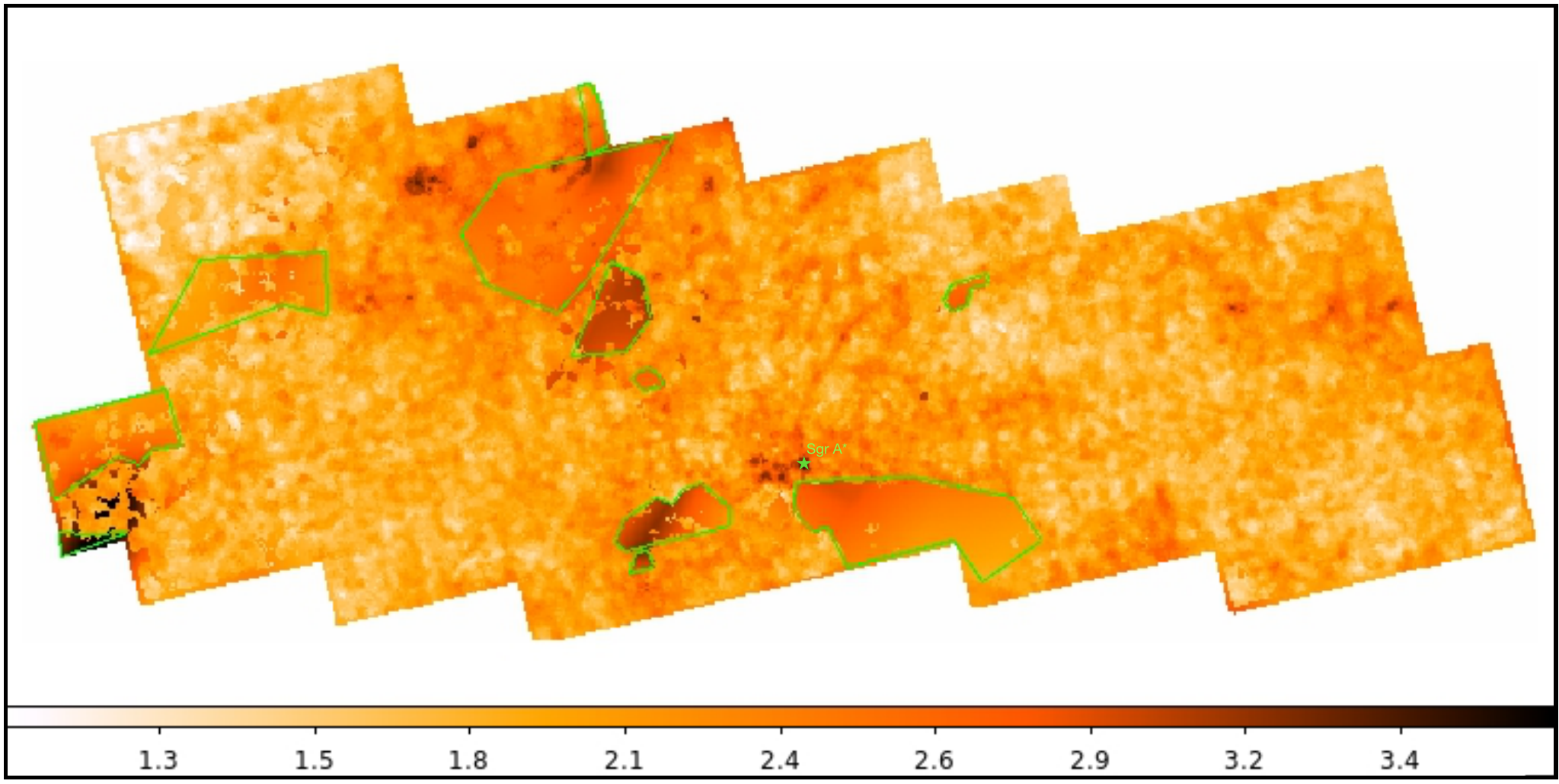}
\caption{\label{Fig:ext_map} Extinction map $A_{K_{s}}$ from the central $416$ arcmin$^2$ ($\sim$1 kpc$^2$) used for the extinction correction. The extinction map is computed by converting the extinction map of \citet{Dong:2011ff} into $K_{s}$ magnitude. We apply masks to some regions (green polygons). The median value of $A_{K_{s}}$ is $1.9$. The spatial resolution is $\sim9.2$ arcsec. The green star shows the position of Sgr\,A*.}
\end{figure*}

We do not correct the effect of extinction on the density of the detected sources because we exclude highly extincted regions. Furthermore, in fields with moderate extinction, differential extinction is $\leq1$ mag. Since our detection limit is about $K_{s}\sim18-19$ in all fields, but our completeness cut off (due to data quality and crowding) is $K_{s}=16$, these variations in extinction are not considered to have any significant impact on our results. We correct the magnitudes of the individual stars using the extinction map of \citet{Dong:2011ff}. These latter authors created the extinction map using $F187N$ and $F190N$
filters\footnote{We assume the following values of the
effective wavelengths: $\lambda_{F187N}=1.8745$ $\mu$m
and $\lambda_{F190N}=1.9003$ $\mu$m.} from the Hubble
Space Telescope/Near-Infrared Camera and Multi-Object
Spectrometer (HST/NICMOS). In order to convert into $K_{s}$ magnitude, we assume that the extinction curve in the NIR can be described by a power law \citep[e.g.,][]{Nishiyama:2008qa,Fritz:2011fk} and we use the extinction index obtained by \citet{2018A&A...610A..83N} $\alpha=2.30\pm0.08$. Figure
\,\ref{Fig:ext_map} shows the extinction map of
\citet{Dong:2011ff} converted into $K_{s}$ magnitude. We
compute the intrinsic $K_{s, extc}$ magnitude for
individual stars using the obtained extinction map.
Figure\,\ref{Fig:densitymaps} shows the extinction-corrected density map for stars with $9.0\leq K_{s,
extc}\leq14.0$. The shape of the FOV of the density map is due to the FOV of the extinction map. Because of dark foreground clouds, the source density in the map of
\citet{Dong:2011ff} is very low and the extinction was averaged over large regions. The latter can be easily identified as large patches in the extinction map that hardly show any variation. We mask those regions by hand because they will be heavily contaminated by foreground stars (see green polygons in Fig.\,\ref{Fig:ext_map}). The same stars are not necessarily used in the uncorrected density map (without applying extinction correction) and in the extinction-corrected density map because the extinction correction may move stars in and out of the faintest considered bin. In Appendix\,\ref{sec:err_ext}, we provide further tests to study the systematic uncertainty derived by the extinction-correction, such as considering a less steep value for the extinction index $\alpha$.

\subsection{Selection of the magnitude ranges for the analyzed stars \label{sec:selection}}
In order to test whether the composition of the stellar population in the MWNSC changes with distance from Sgr\,A*, we study the density maps in the following brightness ranges (in parentheses for extinction-corrected magnitudes):
\begin{enumerate}[(a)]
\item RC stars and bright giants in the ranges: $11.0 \leq K_{s}\leq16.0$ ($9.0 \leq K_{s, extc}\leq14.0$);
\item only bright giants in the ranges: $11.0\leq K_{s}\leq14.0$ ($9.0 \leq K_{s, extc}\leq12.0$);
\item RC stars in the ranges: $14.5 \leq K_{s}\leq16.0$ ($12.5 \leq K_{s, extc}\leq14.0$).
\end{enumerate}
We selected the magnitude ranges of RC and bright stars so as to avoid mixing different kinds of stars. The magnitude ranges in the uncorrected density maps are selected in order to consider roughly the same number of stars as in the extinction-corrected density maps. We study the brightness ranges (a) in order to compare with previous studies. Although we do not need any precision photometry, we select the brightest magnitude in the range of the stars in order to avoid saturation effects in our sample. Furthermore, in Appendix\,\ref{sec:saturation} we explore the results by considering another brightest magnitude limit.


\section{Structure of the nuclear stellar cluster\label{sec:structure_nsc}}
Our main goal is to study the structure of the MWNSC. Since the MWNSC is not isolated, when we compute the star density toward the GC we have to take into account several superposed components: NSD, GB, and GD (in decreasing order of star density). The GD contributes a negligible contamination within the central $\sim10$ pc (see the possible contamination by stars in the GD in Appendix\,\ref{sec:foreground}). The primary source of systematic uncertainty is due to the NSD, and the secondary one is due to the GB. We apply the term "background" to both foreground and background contributions that contaminate our sample. We use the stellar density map of the central $6^{\circ}\times2^{\circ}$ of our Galaxy from \citet{Nishiyama:2013uq} to model the GB and NSD in order to subtract them from the data and isolate the NSC in the stellar density maps (see left panel in Fig.\,\ref{Fig:nsdmodel}). Finally, we fit S\'ersic functions to the stellar density maps of the MWNSC.
Furthermore, we use MIR Spitzer images to constrain the shape of NSC and NSD, following up the work of \citet{Schodel:2014fk}.

\subsection{Galactic bulge \label{sec:Bulge}}
We model the GB as a triaxial ellipsoidal bar with a
centrally peaked volume emissivity $\rho$
\citep{Dwek:1995uq,1997ApJ...477..163S,1998ApJ...492..495F}.
We assume the best GB parameters computed by
\citet[][see Table 4 in their paper and
Table\,\ref{Tab:NSDparameters} in the present paper]{Launhardt:2002nx}. These latter authors found that the best fit for the surface brightness profile of the GB is the one that represents the volume emissivity by an exponential model:
\begin{equation}
\rho(r) \propto
 exp(-R_{s})
.\end{equation}
They selected the shape of the bar as a "generalized ellipsoid" \citep{1990MNRAS.245..130A,1998ApJ...492..495F}, where $R_{s}$ is the effective radius \citep[see equations 5 and 6 in][]{Launhardt:2002nx}.

\begin{figure}[!htb]
\includegraphics[width=\columnwidth]{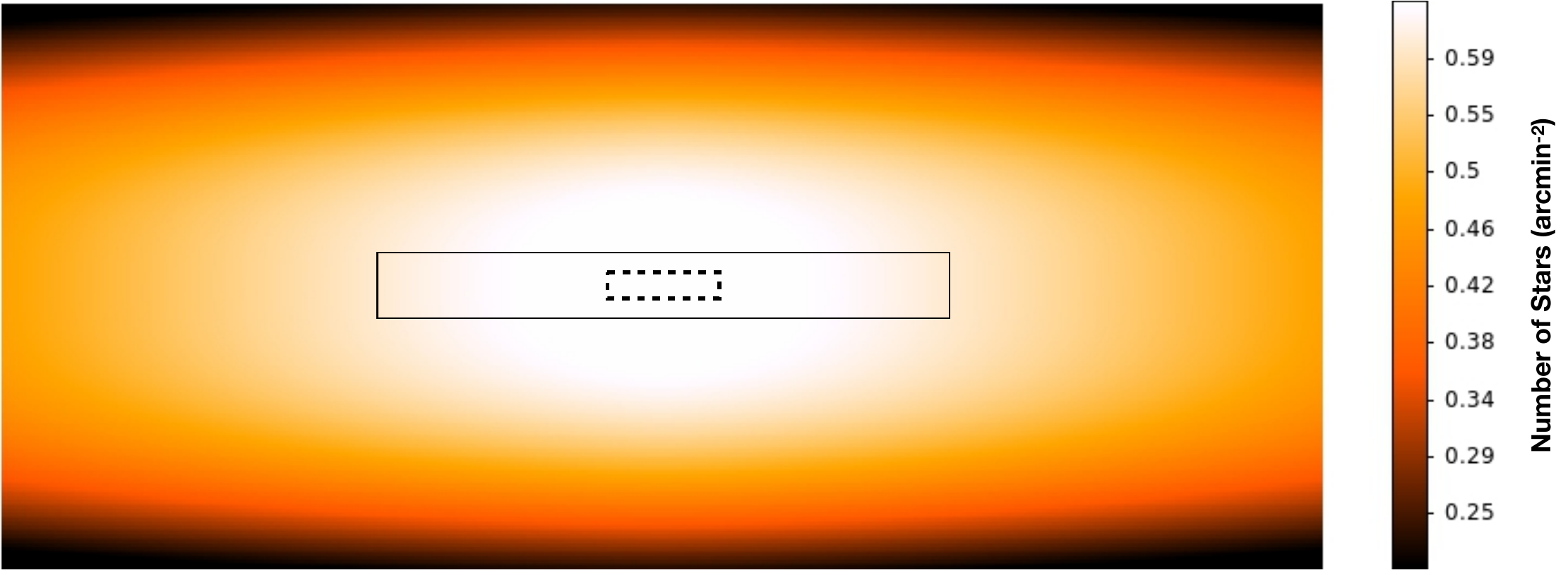}
\caption{\label{Fig:GB} Triaxial ellipsoidal bar model with
a centrally peaked volume emissivity $\rho$ for the GB of
the central $6^{\circ}\times2^{\circ}$ ($\sim$ 864 pc $\times$ 288 pc) of our Galaxy. The inner black dotted rectangle indicates the FOV of our data ($\sim$ 86 pc $\times$ 20.2 pc) and the outer black rectangle indicates the region occupied by the NSD ($\sim$ 460 pc $\times$ 40 pc). Sgr\,A* is located in the middle of the image. Galactic north is up and Galactic
east is to the left. The GP runs horizontally through the
center.}
\end{figure}

The intensity of light from the bulge region in the $\{l,b\}$ direction is given by an integral of the volume emissivity $\rho$ along the line of sight $s$ \citep[for more details see equation \,1 of][]{Dwek:1995uq}:
\begin{equation}
I(l,b)=
2\cdot N_{0}\cdot\int_{0}^{\infty}exp(-R_{s})\cdot ds
.\end{equation}

We apply the previous equation to compute the projected stellar
number density of the GB. The parameter $N_{0}$ is a normalization
constant given in units of stellar number density and is determined
by fitting the model $I(l,b)$ to the stellar number density map
from \citet{Nishiyama:2013uq}. We mask a region of $\sim500$ $\times$ $72$ pc$^{2}$ centered on Sgr\,A* occupied by the NSD
\citep{Launhardt:2002nx} and some dark clouds to fit the GB. We assume that its central stellar density can be derived from its
outer parts. Figure \,\ref{Fig:nsdmodel} shows the GB model that
fits for the stellar density map. Table\,\ref{Tab:NSDparameters}
summarizes the parameters that we assume in the model and the
value of $N_{0}$ that we obtain.

\subsection{Nuclear stellar disk \label{sec:NSD}}
\begin{figure*}[!htb]
\includegraphics[width=\textwidth]{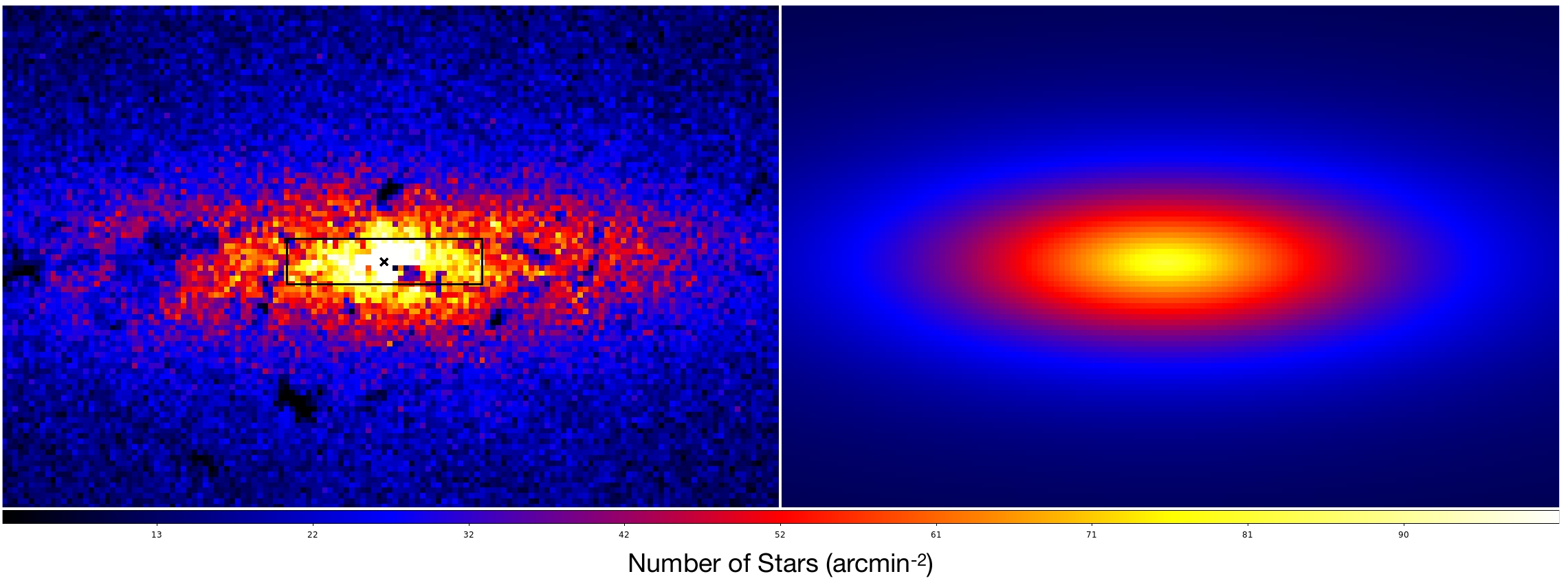}
\caption{\label{Fig:nsdmodel} Map of the stellar density in the central 341\,pc $\times$ 219\,pc of our Galaxy. Both panels use the same color linear scales. \textit{Left panel}: Stellar number density map from \citet{Nishiyama:2013uq}. The black rectangle indicates the FOV of our data ($\sim$ $86$ pc $\times$ $20.2$ pc). Sgr\,A* is located in the position of the black cross. \textit{Right panel}: S\'ersic plus triaxial ellipsoidal fit to the data after masking the central 10 pc occupied by the NSC and regions with low density.}
\end{figure*}

In this section, we analyze the NSD, the component with the overall highest star density and smallest angular scale. We model the NSD using the stellar density map of the central $6^{\circ}\times2^{\circ}$ of our Galaxy of \citet{Nishiyama:2013uq}.
We mask regions with low density (see dark clouds in the right-hand image in Fig. \,\ref{Fig:nsdmodel}) and a region of approximately $10$ pc $\times$ $10$ pc centered on Sgr\,A* occupied by the NSC. The NSD can be modeled by an elliptical model. We use a S\'ersic model \citep{Graham:2001ys} given by

\begin{equation} \label{eq:sersic}
I(x,y)=
I_{e}exp\bigg\{-b_{n}\bigg[\bigg(\frac{p}{R_{e}}\bigg)^{1/n}-1\bigg]\bigg\},
\end{equation}

where $I_{e}$ is the intensity at the effective radius $R_{e}$, which encloses $50\%$ of the light. The factor $b_{n}$ is a function of the S\'ersic index $n$. We use the approximation $b_{n}=1.9992n-0.32 $ given by \citet{Capaccioli:1987vn} for $1<n<10$. We use the modified projected radius $p$ for elliptical models

\begin{equation} \label{eq:projected}
p=
\sqrt{x^2+(y/q)^2},
\end{equation}

 where $x$ and $y$ are the 2D coordinates and $q$ is the ratio between minor and major axis.
We are interested in studying the star count contribution of the NSD, and therefore we convert the $I(x,y)$ into stellar number density in the $\{x,y\}$ position. Finally, we fit the S\'ersic model for the NSD after subtracting the GB model computed in the previous section to the stellar number density map. The best-fit parameters were obtained using a gradient-expansion algorithm to compute a nonlinear least squares fit to the model (IDL CURVEFIT procedure). The value of $\chi^{2}_{reduced}$  is $1.082$. The best-fit parameters are summarized in Table\,\ref{Tab:NSDparameters}.

\begin{table}
\centering
\caption{Galactic bulge and nuclear stellar disk Model 1 parameters using the stellar density map of the central $6^{\circ}\times2^{\circ}$ of our Galaxy of \citet{Nishiyama:2013uq}.}
\label{Tab:NSDparameters}
\begin{tabular}{ll}
\hline
\hline
Galactic bulge \\
\hline
Parameters & Value\\
\hline
Sun dist. from GC $R_{0}^{\mathrm{a}}$ & 8.5 kpc \\
Bar X scale length $a_{x}^{\mathrm{a}}$ & 1.1 kpc  \\
Bar Y scale length $a_{y}^{\mathrm{a}}$ & 0.36 kpc  \\
Bar Z scale length $a_{z}^{\mathrm{a}}$ & 0.22 kpc  \\
Bar face-on parameter $C_{\bot}^{\mathrm{a}}$ & 1.6  \\
Bar edge-on parameter $C_{\parallel}^{\mathrm{a}}$ & 3.2  \\
Normalization constant $N_{0}$  & ($9.90\pm0.02$) stars/arcmin$^{2}$  \\
\hline
\hline
Nuclear stellar disk\\
\hline
Parameters & Value\\
\hline
Stellar number density at $R_{e}$ & ($19.84\pm0.16$) stars/arcmin$^{2}$  \\
Flattening $q$  & ($0.338\pm0.002$) \\
S\'ersic index $n$  & ($0.782\pm0.008$) \\
Effective radius $R_{e}$  & ($116.2\pm0.6$)pc \\

\hline
\end{tabular}
\begin{list}{}{}
\item[$^{\mathrm{a}}$] Fixed input parameter from \citet{Launhardt:2002nx}.
\end{list}
 \end{table}

 Although we assume a larger value for the distance of the GC to compute the GB, if we assume a more updated value of 8\,kpc \citep[see,
e.g.,][]{Genzel:2010fk,Meyer:2012fk,2018A&A...615L..15G} the differences between the parameters obtained using both values are negligible ($<0.001\%$).


\subsection{S\'ersic models\label{sec:sersicmodels}}
\subsubsection{NSC aligned with respect the GP}
\begin{figure}[!htb]
\includegraphics[width=\columnwidth]{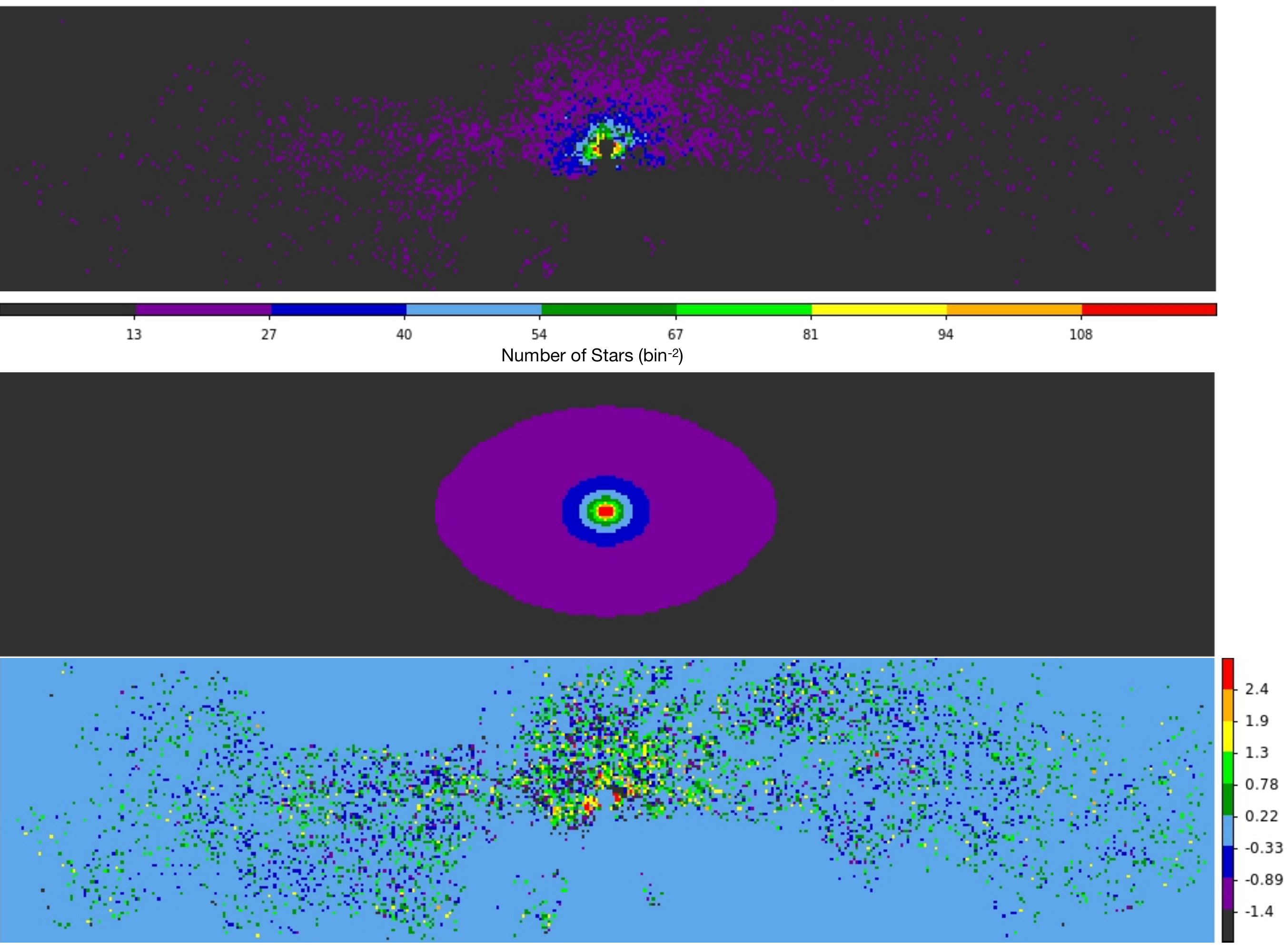}
\caption{\label{Fig:9_14sersicmodel} Comparison between the stellar density map of the central $86$ pc $\times$ $20.2$ pc of the Galaxy and the model. \textit{Upper}: Extinction-corrected stellar density map for stars with $9.0\leq K_{s, extc}\leq14.0$. The dark regions are masked in the fit. \textit{Middle}: S\'ersic model for the NSC plus the background model (ID 10 in Table\,\ref{tab:anexosersicfits}).  \textit{Lower}: Residual image given by the difference between the upper and middle panels divided by the uncertainty map of the density map. The color bar shown for the lower panel is in units of the standard deviation. The other two panels use the same linear color scale. Galactic north is up and Galactic east is to the left.}
\end{figure}

The S\'ersic model gives a good description of the large-scale structure of the NSC \citep{Schodel:2014fk}. We apply it to the extinction-corrected density maps computed for the different magnitude ranges of the stars (see section\,\ref{sec:selection}) and use the following steps to fit the models:

\begin{enumerate}
  \item We apply a mask to take into account the variable completeness across the FOV due to the different observing conditions, extinction, and crowding, as we saw in section\,\ref{sec:crowding}. We study the systematic errors caused by the selection of the masks in Appendix\,\ref{sec:masks}.
  \item To avoid biased results in the S\'ersic model, which has a central peak, we mask the inner $0.6$ pc. This is due to the fact that in our study, the number counts are dominated by bright stars with $K_{s}<16.0$ that show a flat profile, different from the cusp profile, which is shown by faint stars with $K_{s}>16.0$ \citep[see][]{Buchholz:2009fk,Bartko:2010fk,2012ApJ...744...24Y,Do:2013fk,2018A&A...609A..26G,2018A&A...609A..27S}.
  \item We fit a S\'ersic NSC fit to the stellar density maps for the different magnitude ranges, including the models computed for the GB and the NSD in Sects\,\ref{sec:Bulge} and \,\ref{sec:NSD}, respectively. Only the stellar number density at effective radius for the NSD is scaled to the maps, keeping the ratio GB/NSD constant.
  \item We fix the tilt angle between the NSC with respect to the GP to $0$ degrees. In the following section, we leave this angle free.
\end{enumerate}
We found the best-fit solution by using a gradient-expansion algorithm to compute a nonlinear least squares
fit to the model with five or six parameters (IDL
MPCURVEFIT procedure), depending on whether the
inclination angle between NSC and GP is fixed or free,
respectively (see the following section). We explored the
systematic effects of the different choices of masks and
other parameters by changing their values and fitting the
data repeatedly. The resulting best-fit values are listed
in Table\,\ref{tab:anexosersicfits}.
Figure\,\ref{Fig:9_14sersicmodel} shows the comparison
between the density map for stars with $9.0\leq K_{s,
extc}\leq14.0$ (upper panel) and the S\'ersic model plus
the background model (middle panel) that we obtain in the
fit (ID 10 in Table\,\ref{tab:anexosersicfits}). The lower panel shows the residual image given by the difference between the density map and the model divided by the uncertainty map of the density map. The uncertainty map is created by taking the square root of the values in the stellar density map. The residual map is not completely homogeneous, but the differences along the FOV are $<2.5\sigma$ and are concentrated in the inner 1-2 pc, which is probably related to the fact that we have mixed different data sets. These differences are smoothed and the averages of the residuals are smaller when the tilt angle between the NSC and the GP is free in the fits, as we can see in Fig.\,\ref{Fig:residual}.

\begin{figure}[!htb]
\includegraphics[width=\columnwidth]{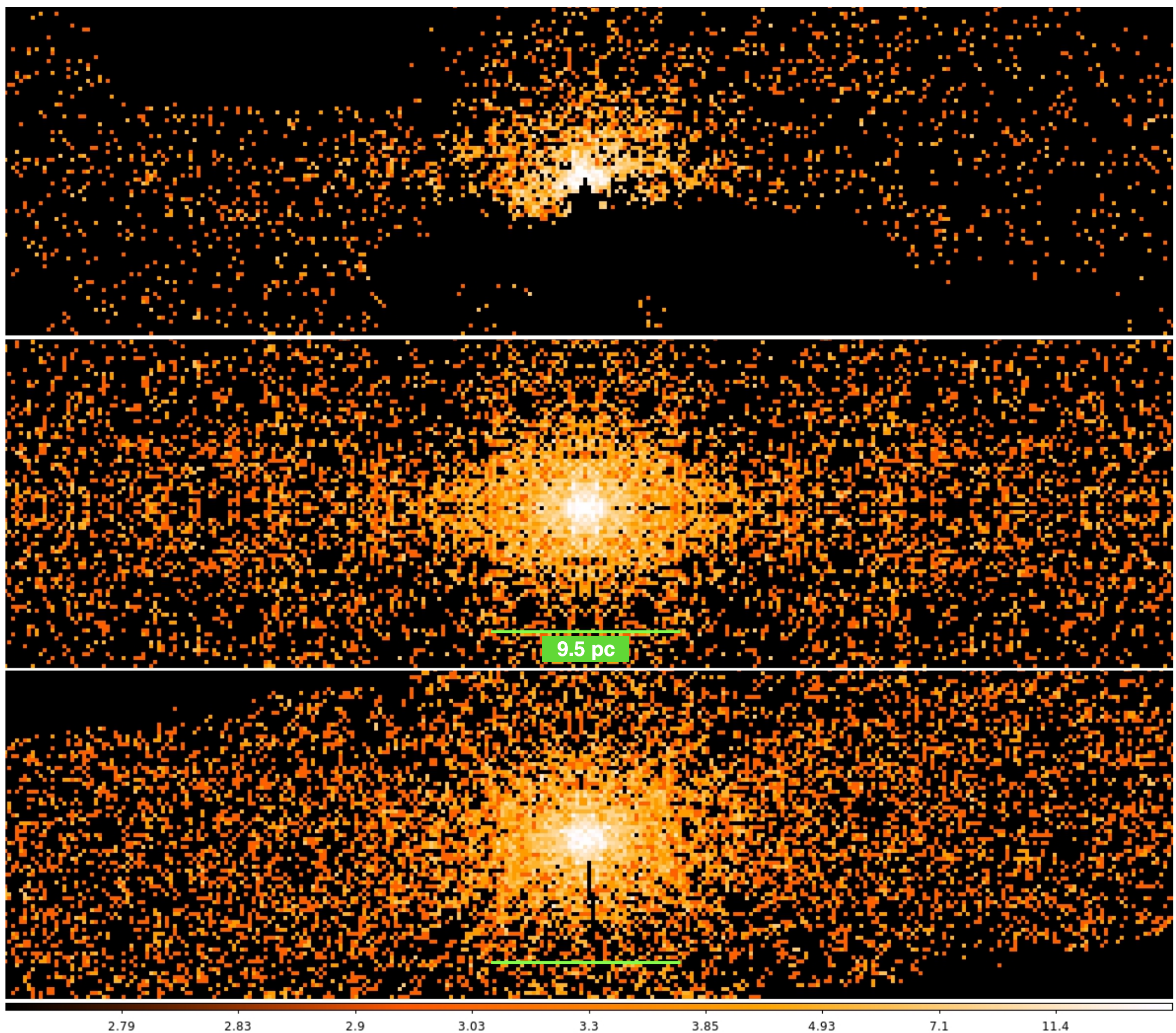}
\caption{\label{Fig:sym_912} Comparison between the stellar density map of the central
$58$ pc $\times$ $16$ pc of the Galaxy and the symmetrized images. \textit{Upper}: Extinction-corrected density map for stars with $9.0\leq K_{s, extc}\leq12$. The mask applied is
shown (dark regions). \textit{Middle}: Symmetrized image is obtained from taking the
median of the pixels from the four quadrants by assuming symmetry of the cluster with
respect to the GP and with respect to the Galactic north--south axis through Sgr\,A*.
\textit{Lower}: Symmetrized image obtained by assuming symmetry of the cluster with
respect to tilted axes through Sgr\,A* with an angle of $\sim -5.6$ degrees. The three panels use the same color scale. Galactic north is up and Galactic east is
to the left. }
\end{figure}

As a second approach to the problem, we symmetrize the cluster with respect
to the GP and with respect to the Galactic north--south axis through Sgr\,A*. We
obtain the symmetrized image by replacing each pixel in each quadrant with the
median of the corresponding pixels in the four image quadrants and ignoring the
masked areas. The pixels along the vertical and horizontal symmetry axes are not
averaged. The uncertainty for each pixel is computed by taking the standard error
of the mean. Figure\,\ref{Fig:sym_912} shows the comparison between the stellar density map (upper panel) and the symmetrized image (middle panel) for stars with $9.0\leq K_{s, extc}\leq12.0$ (ID 7 for Table\,\ref{tab:anexosersicfits}). We fit the S\'ersic model to the symmetrized images corresponding to the different magnitude ranges.

The resulting best-fit values are listed in
Table\,\ref{tab:anexosersicfits}.
Table\,\ref{tab:sersicfits} shows the final parameters that
we obtain taking the mean value of the best-fit parameters
from Table\,\ref{tab:anexosersicfits}. The uncertainties
are the standard deviations of all the best-fit parameters.
We add the statistical errors quadratically to the final
uncertainties.

\begin{table*}[!htb]
\centering
\caption{Best-fit model parameters for S\'ersic fits to the stellar number density maps. The MWNSC is aligned with respect to GP.}
\label{tab:sersicfits}
\begin{tabular}{l l l l l l l l l l}
\hline
\hline
ID  & Mag. range  & $N^{b}_{e,nsd}$ & $N^{c}_{e,nsc}$ & $q^{d}$ & $n^{e}$ & $R^{f}_{e}$ \\
& ($K_{s, extc}$)& (stars/bin$^{2,a}$) & (stars/bin$^{2,a}$) & & & (pc) \\
\hline
1 & $12.5-14$ & $2.45\pm0.10$ & $3.94\pm0.68$ & $0.85\pm0.05$ & $2.59\pm0.27$ & $5.66\pm0.32$ \\
2 & $9-12$ & $0.55\pm0.09$ & $1.16\pm0.26$ & $0.60\pm0.04$ & $2.28\pm0.47$ & $4.57\pm0.78$ \\
3 & $9-14$ & $3.22\pm0.19$ & $6.55\pm1.08$ & $0.78\pm0.02$ & $2.02\pm0.46$ & $5.38\pm0.21$ \\

\hline
\end{tabular}
\tablefoot{
\tablefoottext{a}{The bin size in the density maps is $5"\times5"$. }\\
\tablefoottext{b}{Stellar number density for the NSD at its effective radius (see value in Table\,\ref{Tab:NSDparameters}). }\\
\tablefoottext{c}{Stellar number density for the NSC at the effective radius $R_{e}$. }\\
\tablefoottext{d}{The flattening $q$ is equal to the minor axis divided by the major axis. }\\
\tablefoottext{e}{$n$ is the S\'ersic index. }\\
\tablefoottext{f}{$R_{e}$ is the effective radius for the NSC. }
}
\end{table*}


\subsubsection{Tilt angle\label{sec:rotation_angle}}
\begin{figure*}[!htb]
\includegraphics[width=\textwidth]{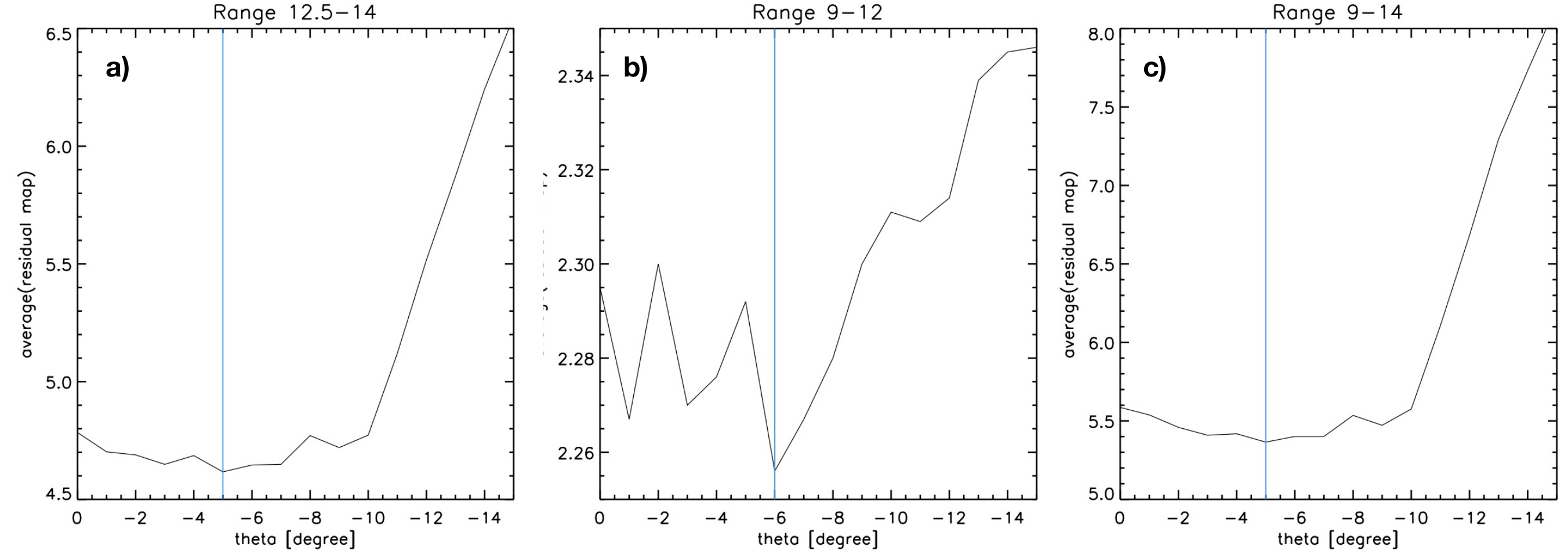}
\caption{\label{Fig:residual} Average of the residuals in the difference image between the original and the
symmetrized image of the NSC vs. the value of the tilt angle $\theta$ used to
symmetrize the image. \textit{a)} Average of the residuals for stars with $12.5<
K_{s,ext}<14.0$. We find a minimum at $\theta=-5^{\circ}$ that is not very significant.
\textit{b)} Average of the residuals for stars with $9.0< K_{s,ext}<12.0$. We find a
minimum for $\theta=-6^{\circ}$. \textit{c)} Average of the residuals for stars with
$9.0< K_{s,ext}<14.0$. In this case, we obtain similar results for stars with $12.5<
K_{s,ext}<14.0$. For the three cases, the residuals are similar from
$\theta=0^{\circ}$ to $\theta=-10^{\circ}$. The blue lines indicate the value of
$\theta$ that minimizes the residuals.}
\end{figure*}

In this section, we explore the effect of leaving the tilt angle
between the NSC and the GP free in the fits without assuming any
symmetry of the cluster a priori. First of all, we create
different symmetrized maps varying the tilt angle $\theta$ from
$\theta=0^{\circ}$ (NSC aligned to the GP) to $\theta=-15^{\circ}$
(NSC is symmetric with respect to an axis rotated $\theta$ with
respect to the GP and with respect to an axis perpendicular to GP
through Sg\,A*). We compute the residual images given by the
difference between the stellar density map and the symmetrized
image. Figure\,\ref{Fig:residual} shows the average of the
residual maps given by the root mean square of all values for the
different magnitude ranges. We can see that there is an
improvement in the residual for a tilted symmetrized cluster
compared to a nontilted one. For bright stars, we find a minimum
value around $\theta=-6^{\circ}$ (middle panel in
Fig.\,\ref{Fig:residual}). For RC stars, we find a minimum around
$\theta=-5^{\circ}$ (left panel in the figure). We obtain similar
results for stars with $9.0\leq K_{s,ext}\leq14.0$ (right panel in
the figure). For all cases, the average of the residuals are very
similar from $\theta=0^{\circ}$ to $\theta=-10^{\circ}$. We also
performed a fit of NSC + NSD on the untilted and unsymmetrized
map, this time introducing the tilt angle as an additional free
parameter. As we can see in Fig.\,\ref{Fig:residual}, there are
local minima for the value of $\theta$, which encouraged us to
explore the fits considering different initial values for the
angle (see Appendix\,\ref{sec:Initial_rotation}).

The resulting best-fit values are listed in
Table\,\ref{tab:anexosersicrotatefits}. The lower panel of figure\,\ref{Fig:sym_912} shows the
symmetrized image corresponding to ID 7 in
Table\,\ref{tab:anexosersicrotatefits}.
Table\,\ref{tab:rotatesersicfits} shows the final results
that we obtain taking the mean value of the best-fit
parameters from Table\,\ref{tab:anexosersicrotatefits} and
the uncertainties are the standard deviations of the
systematic errors. The statistical errors for the parameters
are negligible. In the case of $\theta$, we have to
also include the systematic errors associated to the selection of
different initial $\theta$ (see
Appendix\,\ref{sec:Initial_rotation}). In this case, we also assume the statistical errors. The value of $\theta$
obtained for bright stars is very consistent with the value
that minimizes the residuals. For the other stars, the
values are inside the range of the angles that minimize the
residuals.

\begin{table*}[!htb]
\centering
\caption{Best-fit model parameters for S\'ersic fits to the stellar number density maps. The tilt angle between the NSC and the GP is a free parameter in the fits.}
\label{tab:rotatesersicfits}
\begin{tabular}{l l l l l l l l}
\hline
\hline
ID  & Mag. range  & $\theta^{a}$  & $N_{nsd}$ & $N_{nsc}$ & $q$ & $n$ & $R_{e}$ \\
& ($K_{s, extc}$)& (degree) & (stars/bin$^{2}$) & (stars/bin$^{2}$) & & & (pc) \\
\hline

1 & $12.5-14$ & $-8.51\pm3.15_{sys}\pm0.66_{stat}$ & $2.43\pm0.10$ & $3.38\pm0.61$ & $0.83\pm0.04$ & $3.01\pm0.39$ & $6.19\pm0.68$ \\
2 & $9-12$ & $-5.60\pm1.28_{sys}\pm0.57_{stat}$ & $0.55\pm0.08$ & $1.24\pm0.25$ & $0.64\pm0.10$ & $1.93\pm0.50$ & $4.31\pm0.52$ \\
3 & $9-14$ & $-6.98\pm1.25_{sys}\pm0.40_{stat}$ & $3.21\pm0.20$ & $6.63\pm1.27$ & $0.75\pm0.01$ & $2.26\pm0.21$ & $5.36\pm0.36$ \\

\hline
\end{tabular}
\tablefoot{
\tablefoottext{a}{The tilt angle of NSC in degrees. Positive angle is clockwise with respect to GP.}
}

\end{table*}

\subsection{Best-fit NSC parameters from MIR imaging \label{sec:mir}}
To obtain an independent constraint on the projected structure of the NSD plus NSC, we
also fit our models to IRAC/Spitzer $4.5 \mu$m images (see Fig.\,\ref{Fig:spitzer}).
We use the same data and technique as outlined in \citet{Schodel:2014fk}. However,
while the latter authors used images no larger than  $200$ pc $\times$ $200$ pc in order to
avoid the difficulties of the spatial change in the background contributed by the GB
and due to large dark clouds, here we fit the NSD to the full images
($\sim$ $300$ pc $\times$ $250$ pc) as delivered by the survey of
\citet{Stolovy:2006fk} and include our GB model to take the influence of the latter
into account. All large dark clouds were masked. By using the full extent of the
images from the Spitzer survey we expect to obtain better constraints on the NSD with
its large scale length. We repeat the double S\'ersic NSC plus NSD fit of
\citet{Schodel:2014fk}, including also our GB emission model based on
\citet{Launhardt:2002nx} and leaving the tilt angle free.
Tables\,\ref{Tab:nsd_mir_img} and\,\ref{Tab:nsc_mir_img} show the final results
obtained for the NSD (NSD Model 2) and NSC, respectively. Our results are very similar
to what was obtained by \citet{Schodel:2014fk}, but should be better constrained
because of the larger FOV and the use of a more accurate GB model. Due to the low
resolution of the data, we can only approximately compare the results with the values
obtained for stars with $9.0\leq K_{s, extc}\leq12.0$ (ID 2 in
Table\,\ref{tab:rotatesersicfits}). All the parameters that we obtain in the fits are
consistent with the final results at the $1\sigma$ level.

\begin{table}
\centering
\caption{Nuclear stellar disk Model 2  using MIR imaging.}
\label{Tab:nsd_mir_img}
\begin{tabular}{ll}
\hline
\hline
Nuclear stellar disk\\
\hline
Parameter & Value\\
\hline
$I_{e}^{\mathrm{a}}$ & ($0.279\pm0.003$) mJy arcsec$^{-2}$\\
$q$ & ($0.372\pm0.005$) \\
$n$  & ($1.09\pm0.03$) \\
$R_{e}$  & ($86.9\pm0.6$)pc \\
\hline
\end{tabular}
\begin{list}{}{}
\item[$^{\mathrm{a}}$] Flux density at $R_{e}$.
\end{list}
 \end{table}

 \begin{table}
 \centering
 \caption{Nuclear stellar cluster model parameters using MIR imaging.}
 \label{Tab:nsc_mir_img}
 \begin{tabular}{ll}
 \hline
 \hline
 Nuclear stellar cluster\\
 \hline
 Parameter & Value\\
 \hline
 $\theta^{\mathrm{b}}$ &  $(-6.9\pm1.7)$ degree \\
 $I_{e}^{\mathrm{a}}$ & ($2.64\pm0.43$) mJy arcsec$^{-2}$\\
 $q$ & ($0.66\pm0.03$) \\
 $n$  & ($1.8\pm0.1$) \\
 $R_{e}$  & ($4.9\pm0.5$)pc \\
 \hline
 \end{tabular}
 \begin{list}{}{}
 \item[$^{\mathrm{a}}$] Flux density at $R_{e}$.
 \item[$^{\mathrm{b}}$] Tilt angle defined positive in the direction east of Galactic north.
 \end{list}
  \end{table}

\section{Discussion \label{sec:discussion}}

\subsection{Overall properties of the MWNSC}

The present paper fills some of the gaps that may be present in previous studies. We address the study of the MWNSC properties using two independent methods: an improved analysis of IRAC/Spitzer MIR images, which are less affected by interstellar extinction than NIR images \citep{Nishiyama:2009oj,Fritz:2011fk}, and analyzing star counts and higher-resolution NIR images of NACO and HAWK-I. We aim to build a bridge between the last two comparable studies of \citet{Schodel:2014fk} and \cite{2016ApJ...821...44F} and improve on their shortcomings, as described in the following.

Table\,\ref{tab:comparison_previous} shows a comparison between the results obtained here and in previous studies. The values inferred in this and in previous studies agree within their $1-2\sigma$ uncertainties, indicating that none of them suffer any major bias. Due to the higher angular resolution achieved in our study, we are able to constrain the range of the stars to investigate the distribution of giants in the RC and at brighter magnitudes, separately. Due to its generally very well defined brightness and mass, the RC is a very well suited tracer to studying stellar structures older than $\sim2$ Gyr. The RC stars are the most abundant stars at $K_{s} \leq17$ in the GC and are therefore an important tracer population, but these were not isolated in previous studies because of their low angular resolution, which meant that brighter stars dominated the measurements.

We use the same NIR wavelength and method of star counts as  \cite{2016ApJ...821...44F} but with a more than twice higher angular resolution, which allows us to use the RC as a tracer of structure, limit the influence of crowding, and improve the number statistics. Also, we explicitly model and take into account stellar structures that overlap with the NSC along the line of sight, in particular the Galactic bulge and the nuclear stellar cluster. Finally, \cite{2016ApJ...821...44F} could not clean their sample of foreground stars across their entire field because they did not have two filter measurements for all regions. Here, we can reliably exclude foreground stars.

\citet{Schodel:2014fk} use MIR images and study also the contribution of the GB and NSD. The disadvantages are that the resolution is lower than in our study and they only used the central most parts of the images available to them, which may have biased their work. We repeat their study and improve on it by analyzing the images in their entirety and including the GB model.

\begin{figure*}[!htb]
\includegraphics[width=\textwidth]{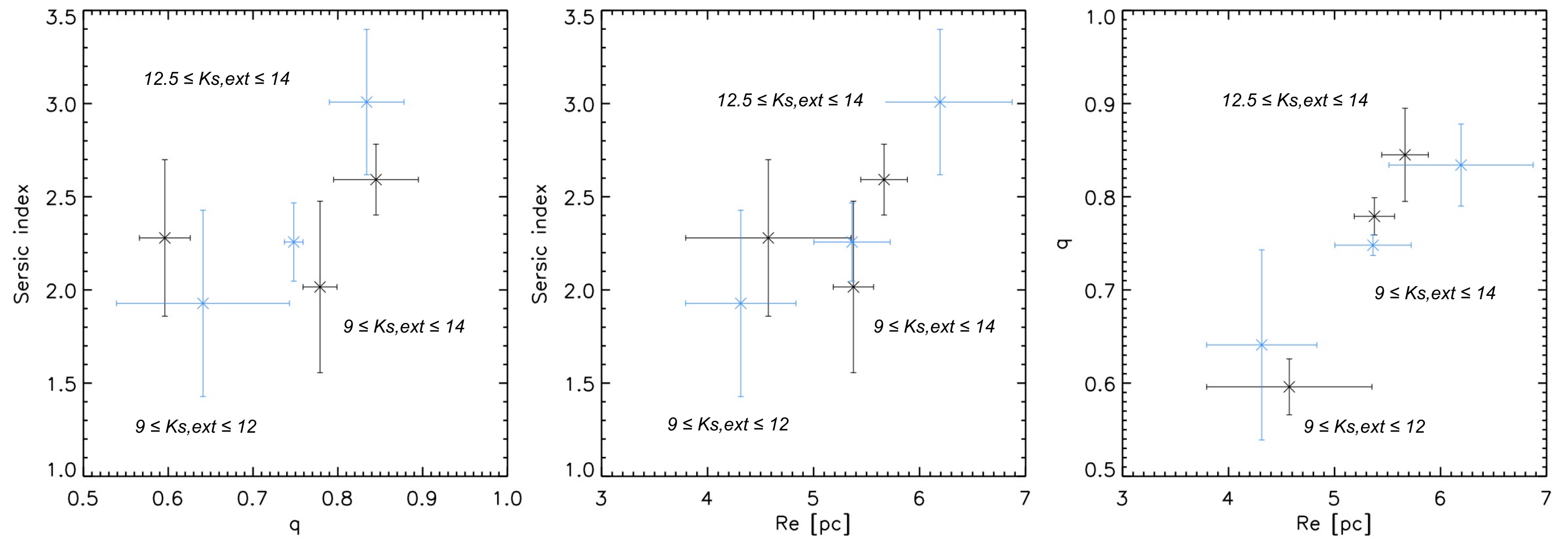}
\caption{\label{Fig:comp_parameters} Comparison between the best-fit NSC parameters for a MWNSC that is assumed to be aligned with the GP (black points) or for the case when a small tip-angle is included in the fits (blue points). The magnitude of the stars is indicated in the panels.}
\end{figure*}

In order to give the final values of the MWNSC parameters,
ID 1 and ID 2 in Table\,\ref{tab:rotatesersicfits} can be
taken as independent measurements because the samples
that we study in each case are different; they are composed of stars
in two different ranges of magnitude. If we add also the
results from MIR imaging (Table\,\ref{Tab:nsc_mir_img}), we
can assume that the final values obtained from our
work are the mean of the three parameters and the
uncertainties are the standard deviations of them. We obtain an
effective radius: $R_{e}=(5.1 \pm1.0)$ pc; an axis
ratio: $q=0.71 \pm0.10$; a tilt angle: $\theta=(-7.0
\pm1.5)$ degrees ; and a S\'ersic index: $n=2.2
\pm0.7$. The ellipticity is $\epsilon=1-q=0.29 \pm0.10$.
The effective radius and the ellipticity of the MWNSC are
very well constrained (see
Fig.\,\ref{Fig:comp_parameters}).
\cite{2016ApJ...821...44F} study the flattening in
different ranges of distance from the SMBH. These latter authors report
$q=0.80$ at distances smaller than $2.3$ pc, but for larger
distances, they obtain larger values of the flattening,
very similar to our results. The reason for these somewhat different
results is probably linked to the fact that they do not take into
account the MWNSC as a separate entity from the GB and NB.
The value of the S\'ersic index is highly dependent on the
distribution in the central parsec.
\cite{2016ApJ...821...44F} consider the central parsec in
the fits in the same way that we do in the present work,
but we mask the central $0.6$ pc to avoid contamination from
young stars and biases results of the fits because of the
flat profile showed by bright stars \citep{2018A&A...609A..26G}. \citet{Schodel:2014fk}
reported a smaller value for the S\'ersic index than
ours. The difference comes from the NIR images, because
they are forced to mask the central parsec because of the low
resolution of the data and the existence of a few
extremely bright sources and a strong diffuse emission from
the mini spiral. It appears that the S\'ersic index is very susceptible to bias and caution is required in its
application. Its precise value also depends strongly on the
assumptions on, and fit of, any structures overlapping with
the NSC.

Both of these latter two studies quote that they may underestimate the systematic
errors in component fitting. We give very robust and reliable uncertainties on the
best-fit NSC parameters, not only by taking into account different potential sources
of systematic error but also by analyzing the MWNSC structure using two completely
different approaches: MIR and NIR imaging. We explore in detail the systematic
uncertainties that can affect the parameters, taking into account different potential
sources of systematic error. We choose to let these uncertainties be reflected in the
error bars of the best-fit parameters in order to give robust and reliable values.

\begin{figure*}[!htb]
\includegraphics[width=\textwidth]{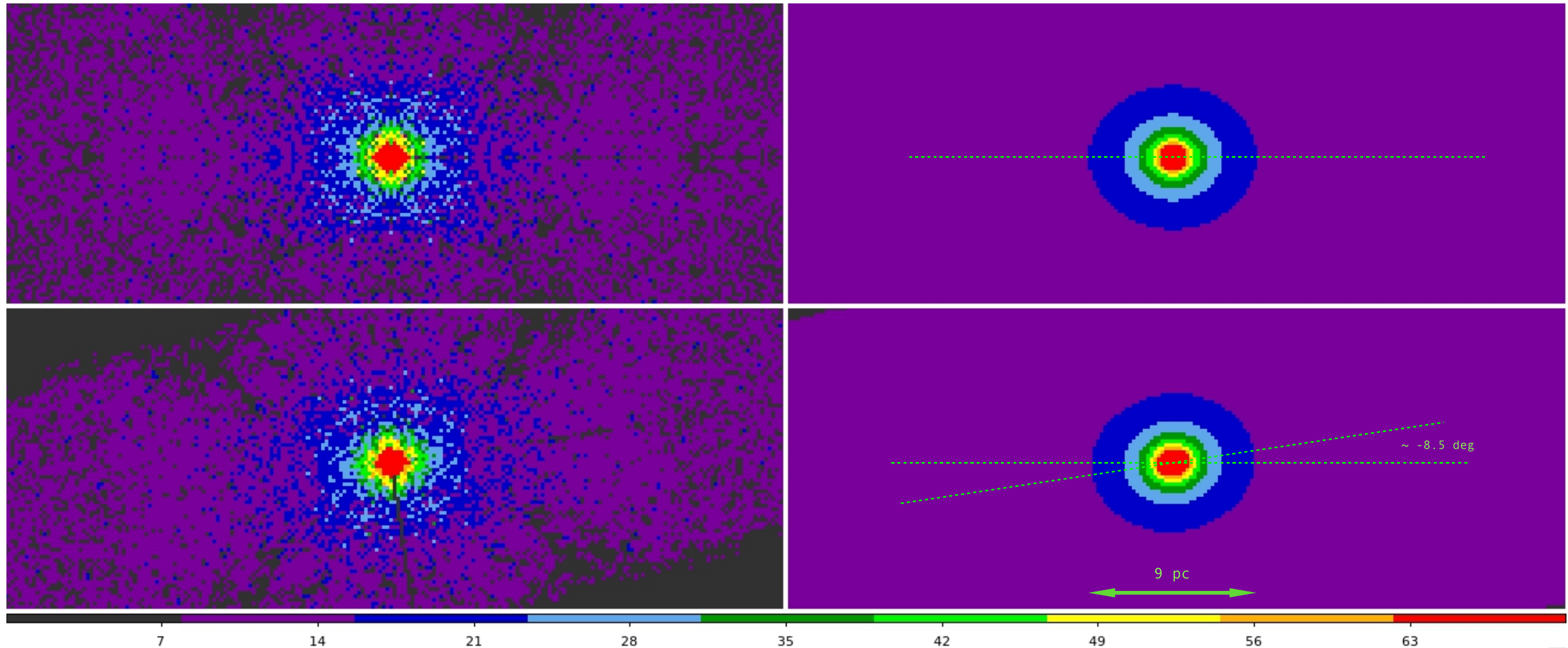}
\caption{\label{Fig:fig12} Zoom-in to compare the symmetrized images of the inner $\sim42$ pc x $16$ pc of the Galaxy for stars with $12.5\leq K_{s, extc}\leq14.0$ obtained by keeping the angle between the NSC and GP equal to zero (upper panels, ID 3 in Table\,\ref{tab:anexosersicfits}) and leaving the tilt angle free in the fits (lower panels, ID 3 in Table\,\ref{tab:anexosersicrotatefits}). Right panels show the S\'ersic model that we obtain from the fits for each case. The color scale is linear and all the images have been scaled identically. Galactic north is up and Galactic east is to the left.}
\end{figure*}

Finally, in addition to assuming that the MWNSC is aligned with respect to the GP, as
in previous studies (upper panels in Fig.\,\ref{Fig:fig12}), we leave the tilt angle
between the NSC and the GP to be free in the S\'ersic models (lower panels in
Fig.\,\ref{Fig:fig12}). We obtain novel photometric data that suggest that the MWNSC
is consistent with a tilt of the major axis out of the Galactic plane by up to -10
degrees. This result corroborates the misalignment of the kinematic position angle of
$\sim-9$ degrees found by \citet{Feldmeier:2014kx}, which is somewhat smaller but is
within the uncertainties. \citet{2016ApJ...821...44F}   also determine the orientation
of the major axis in proper motion data, finding that it agrees within 1.2 degrees
with the GP, in contrast to line of sight and star distribution data. This latter
result suggests that a simple rotation of the cluster is not sufficient to explain all
the data. Figure\,\ref{Fig:fig12} shows a comparison between the symmetrized images of
the inner $\sim42$ pc x $16$ pc of the Galaxy for stars with $12.5\leq K_{s,
extc}\leq14.0$ (left panels) and the S\'ersic models (right panels). For upper panels
we assume that the MWNSC is aligned with respect to the GP in the fits (ID 3 in
Table\,\ref{tab:anexosersicfits}). For lower panels, we leave the tilt angle free in
the fits (ID 3 in Table\,\ref{tab:anexosersicrotatefits}).


\begin{table*}[!htb]
\centering
\caption{Comparison with previous studies.}
\label{tab:comparison_previous}
\begin{tabular}{l l l l l l l}
\hline
\hline
Study  & Wavelength  & Angular resolution & Scales & $q$ & $n$ & $R_{e}$ \\
& ($\mu m$)& (arcsec) & (pc$^{2}$) & & & (pc) \\
\hline
Sch\"odel et al. (2014) & $4.5$ & $2$ & $200\times200$ & $0.71\pm0.02$ & $2\pm0.2$ & $4.2\pm0.4$ \\
Fritz et al. (2016) & $2.2$ & $1$ & $80\times80$ & $0.80\pm0.04$ & $1.5\pm0.1$ & $7\pm2$ \\
Present work NIR & $2.2$ & $0.2$  & $86\times20.2$ & $0.75\pm0.01$ & $2.26\pm0.21$ & $5.36\pm0.35$ \\
Present work MIR & $4.5$ & $2$  & $300\times250$ & $0.66\pm0.03$ & $1.8\pm0.1$ & $4.9\pm0.5$ \\
\hline
\end{tabular}
\tablefoot{
\cite{2016ApJ...821...44F} use data with different resolution. We
show the resolution for larger distances (R $>2.72$ pc) where they
used VISTA data. For distances R $\leq0.8$ pc, they use NACO data
with an angular resolution of 0.08". For distances $0.8$
pc $\leq$ R $\leq2.72$ pc they use (HST) WFC3/ IR data with an
angular resolution of 0.15". The present study presents results
for stars with magnitudes in the range $11.0 \leq K_{s}\leq16.0$
(ID 3 in Table\,\ref{tab:rotatesersicfits}). We show the
resolution for larger distances where we use the GALACTICNUCLEUS
survey. For distances R $\leq2$ pc, we use NACO data with an
angular resolution of 0.05". We also present the results from using MIR.
}
\end{table*}

Figure\,\ref{Fig:comp_parameters} shows a comparison between the best-fit parameters for the case when the MWNSC is aligned with the GP and for the case when the tilt angle is left free. The values are consistent with each other inside the uncertainties. Considering also Fig.\,\ref{Fig:residual}, which shows the average of the residuals obtained by the subtraction of the symmetrized image of the MWNSC from the original image, we can see that it is very difficult to constrain the tilt angle of the system, but out-of-plane tilt of up to $-10^{\circ}$ cannot be excluded.

\subsection{Distribution of the stars at the Galactic center}
 The stars in our sample are predominantly giants, mostly
 in the RC and Asymptotic Giant Branch (AGB) with ages of more
 than 5~Gyr \citep{Blum:1996mz,Pfuhl:2011uq}. An
 overabundance of bright stars has been found in the
 Nuclear Bulge (NB) formed by the NSD + the MWNSC
 compared with the stellar population of the GB/Bar
 \citep{Blum:1996mz,Philipp:1999nx}. This result is a
 consequence of different star formation histories of
 both entities since the GB  has not experienced any
 significant amount of recent star formation, contrary
 to the NB \citep[see][]{2018A&A...620A..83N}. The low angular resolution of previous studies made any
 potential differences between the distribution of RC
 stars and brighter stars difficult to study. If
 the ratio between stellar populations of different ages
 changes, the ratio between the number of bright giants
 and giants in the RC will change. For example, adding
 populations of ages $\lesssim 1.5$ Gyr can enhance the
 number of bright giants in the luminosity function.

\begin{figure}[!htb]
\includegraphics[width=\columnwidth]{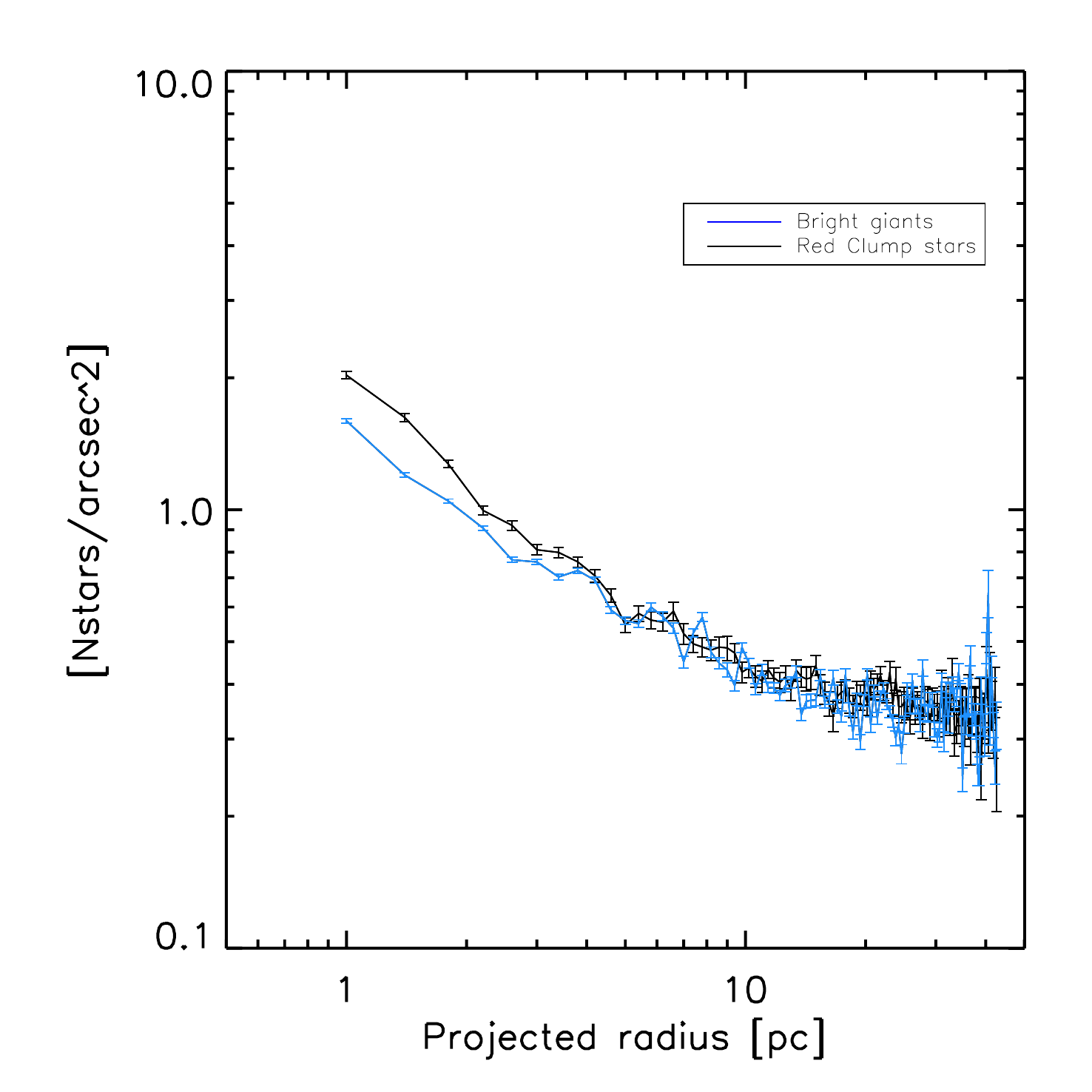}
\caption{\label{Fig:allprofiles_gp_bin10} Comparison between the projected radius profiles computed for different ranges of magnitudes along the GP. Only the data closer than 50" (2pc) to the GP are taken into account. The black line shows the density profile for RC stars ($12.5\leq K_{s, extc}\leq14$) and the blue line indicates the density profile for bright stars ($9.0\leq K_{s, extc}\leq12$). In order to facilitate the comparison between both profiles, the data for the bright stars are scaled using the median ratio of both number counts at distances larger than $10$pc. We mask the inner 0.6 pc around Sgr\,A*.}
\end{figure}

In this work we find slightly different best-fit parameters for the MWNSC, depending on whether we fit our models to the star counts of RC stars or those of brighter giants. The $R_{e}$ for RC stars appears somewhat larger than for bright stars, and the distribution of the latter is more flattened. Furthermore, we study the projected radius profiles for both distributions (see Fig.\,\ref{Fig:allprofiles_gp_bin10}). Although the differences between both distributions are not very significant, they may suggest that the distributions of the stellar populations within the MWNSC are not homogeneous, as a result of distinct star formation histories at different distances from the SMBH. The flat distribution of the bright stars may indicate that some of the stars belong to a star formation event or a dissolved star cluster of a few hundred million years old. However, we are unable to make any firm conclusions because the innermost fields of HAWK-I data that correspond to the MWNSC have a lower S/N and are of lower resolution than further out, and greater resolution is required across the FOV to avoid being dominated by systematic error.

When we fit the S\'ersic model to the MWNSC, we subtract the NSD model from it. There are two important caveats. On the one hand, we use the density map from
\citet{Nishiyama:2013uq} to model the NSD for NIR images.  These latter authors built
the density map for stars significantly brighter than those that we analyzed (stars
with $K_{s, extc}$ brighter than $8.0$ for the central $20'$ and $K_{s, extc}$
brighter than $10.5$ beyond $20'$). Unfortunately, we cannot estimate the effect of
this difference on our results.  However, the fact that we obtain similar results
using NIR and MIR data where we use two different NSD models gives us confidence in
our results. On the other hand, if the stellar populations are different in the
NSC and the NSD, this would also lead to different results for the stars
corresponding to different magnitude ranges. Currently, the dissimilarity in the
stellar
populations has not yet been investigated, but there are indications of differences \citep{2019arXiv191006968N}. In conclusion, the hypothesis of different star formation histories depending on the distance from Sgr\,A* will need to be investigated in a future study using higher angular resolution imaging and spectroscopy.


\subsection{New constraints on the structure of the NSD}

We present two new analyses of the
NSD using completely different data and
methods. For one of these, we use the stellar
number density map from
\citet{Nishiyama:2013uq} of the central
$\sim860$ pc $\times$ $280$ pc of our Galaxy (NSD
Model 1, Table\,\ref{Tab:NSDparameters}) with a bin size of $1'$,
and for the other we derive a density map using IRAC/Spitzer
$4.5 \mu$m images of the central
$300$ pc $\times$ $250$ pc (NSD Model 2,
Table\,\ref{Tab:nsd_mir_img}), with a bin size of $5''-7''$. The
main advantage of using NSD Model 1 is that
we apply a similar procedure to fit the
S\'ersic model to the extinction-corrected
stellar number density maps built from NIR
images using a star count analysis.
Furthermore, the images are larger which
allows us to better constrain the GB model.
The main advantage of NSD Model 2 is that the
images are of higher resolution than the images
from NSD Model 1. Also, the MIR images are
less affected by extinction and were
corrected for it. Taking into account the
effect of extinction is of great importance
in the highly extincted GC region. In
particular, dust in the central molecular
zone appears in projection concentrated
toward a strip between $\pm 30$ pc around the
GP (e.g., Fig. 6 in \citealt{Schodel:2014fk};
see also \citealt{Molinari:2011fk}). Although
the brightness of the individual stars in the
data set that underlies Model 1 has been
corrected for extinction, the overall
systematic effect of the nonrandom
distribution of highly extinguished regions
on this model has not been taken into
account. Such a bias would result in a
systematically thicker NSD model. Similarly,
a gradient of extinction along Galactic
longitude can bias the model toward a larger
effective radius. We therefore believe that
Model 2 is the most accurate one. 

\subsection{Comparison with extragalactic NSCs}

This study is complementary to previous investigations of the structure of the MWNSC.  Having addressed the potential shortcomings of previous studies through the use of sensitive, high-angular-resolution star counts over a large field and careful modeling of the surrounding structures, we have now firmly established that the MWNSC has an effective radius of $R_{e} = 5.1 \pm 1.0$ pc and that it is flattened, with its major axis almost parallel to the Galactic plane. The present study contributes the following two new findings: (1) The major axis of the MWNSC may be tilted out of the Galactic plane by as much as -10 degrees. (2) We find that the giants brighter than the RC show a different distribution to that of the faint stars; in particular, the former display a significantly stronger flattening.

\begin{figure}[!htb]
\includegraphics[width=\columnwidth]{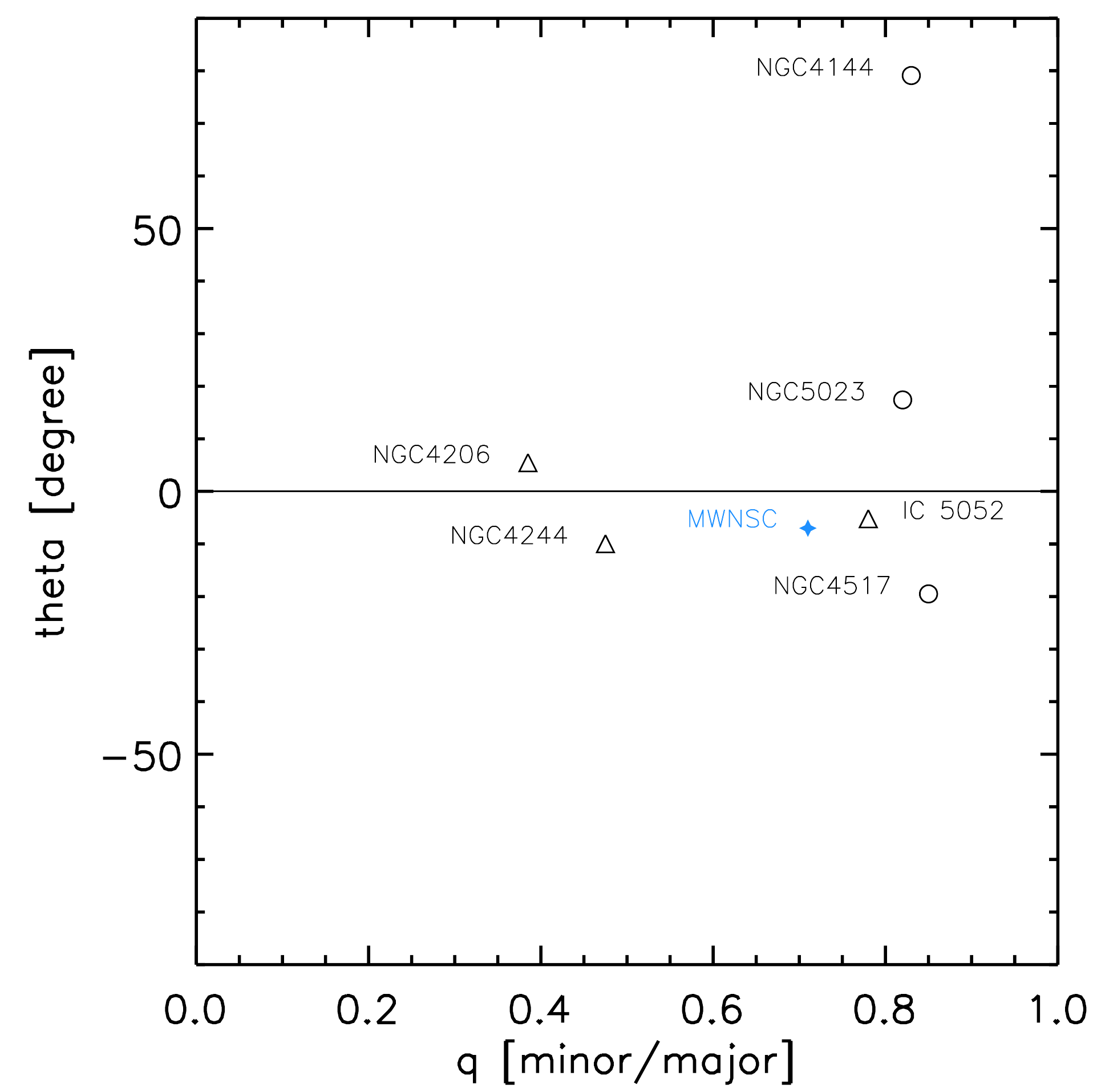}
\caption{\label{fig:tilt_extrag} Axis ratio "q" vs. "theta", the tilt angle between
the NSC major axis and the host galaxy plane for different NSCs found in edge-on
spiral galaxies \citep[data are from Table 2 in][]{Seth:2006uq}. The data have been
averaged for the two different filters for every NSC. The name of the galaxy is
indicated in the panel. The three "multicomponent" nuclear clusters are all aligned
within -10 degrees of the galactic plane of the host galaxy (triangles), similar to
the value of the MWNSC found in the present study (the blue star). All the NSCs are
flattened along the plane of the galaxy, similar to what is observed for the MWNSC.}
\end{figure}

The size and mass of the MWNSC lie well within the distribution of these quantities
for other NSCs in spiral galaxies \citep[e.g.,][]{Georgiev:2014ve,Schodel:2014bn}.
Also, the complex star formation history of the MWNSC is consistent with those of
other NSCs. A range of structural variability has also been found in NSCs in other
spiral galaxies \citep{Georgiev:2014ve}. Flattening along the galactic plane and
potentially small tilt angles between the NSC major axes and the host galaxy planes
have been observed in detail in high-angular-resolution observations of a small number
of nearby spirals \citep{Seth:2006uq,2015AJ....149..170C}.
Figure\,\ref{fig:tilt_extrag} represents the axis ratio versus tilt angle between the
NSC major axis and the galaxy plane for different NSCs in edge-on spiral galaxies
\citep[data are from Table 2 in][]{Seth:2006uq}. We average the results from the two
different filters for every NSC. Three of the clusters in the sample of \citet[][IRAS
09312-3248, NGC 3501, and NGC 4183]{Seth:2006uq} are very compact and
are surrounded by complex emission; we exclude them from Fig.\,\ref{fig:tilt_extrag}
because their values do not come from good fits \citep[][]{Seth:2006uq}. The three
"multicomponent" nuclear clusters that have an elongated disk or ring component and a
spheroidal component are all aligned within -10 degrees of the galactic plane of the
galaxy (see triangles in Fig.\,\ref{fig:tilt_extrag}). We can see that the values of
the axis ratio and the angle for the MWNSC (the blue star in the figure) lie inside
the range of values for NSCs, especially very close to the values for multicomponent
NSCs. \cite{2015AJ....149..170C} also find that the effective radius of NSCs may
change depending on the wavelength used in  observations. This indicates that
populations of different ages may not be fully mixed. This is consistent with our
finding of a different structure of the brightest giants. From stellar evolution
models, we can expect that, on average, the bright giants are younger than the ones in
the Red Clump. Interestingly, \cite{2015AJ....149..170C} also find that NSCs appear rounder at redder wavelengths. Since young stellar populations tend to have bluer colors than older ones and, as mentioned immediately above, bright giants may be younger,  on average, than RC stars; this also agrees with our finding of a rounder MWNSC when we measure the distribution of the RC stars, and a flatter one when we measure the brighter giants.

\subsection{Implications for the stellar cusp around Sgr\,A*\label{sec:cusp}}


The formation of a stellar cusp around a massive BH in a dense and dynamically relaxed
SC is a firm prediction of theoretical stellar dynamics
\citep[e.g.,][]{Lightman:1977ly,Bahcall:1976vn,Freitag:2006fk,Hopman:2006eu}. These
latter authors reach the same conclusion using analytical, Monte Carlo, or N-body
simulations: the stellar number density is described by a power law of the form $\rho
\sim r^{-\gamma}$, where $r$ is the distance from the black hole and $\rho$ is the
stellar density. The GC is a unique case to study the existence of a cusp: it has a
SMBH embedded in a massive NSC and is the closest nucleus of a galaxy. After
controversial observational studies that showed an almost flat, core-like distribution
of stars close to the black hole, \cite{2018A&A...609A..26G} and
\cite{2018A&A...609A..27S} found strong evidence of the existence of the stellar cusp around Sgr\,A*. Thanks to improved methodologies, these latter authors reached the faintest stars studied and tested other possible tracer populations. Their results were corroborated theoretically by \cite{2018A&A...609A..28B}, who evolved a star cluster surrounding a central massive black hole over a Hubble time under the combined influence of two-body relaxation and continuous star formation.


In order to describe the 3D shape of the
cluster in detail, \cite{2018A&A...609A..26G}
and \cite{2018A&A...609A..27S} needed to
convert the measured 2D profile into a 3D
density law and deal with projection effects.
For this reason, they added data at projected
radii $R \ge2$ pc from the literature
\citep{Schodel:2014fk,2016ApJ...821...44F}.
They used the {\it Nuker} model from \cite{Lauer:1995fk} as a
generalization of a broken power law as given
in equation \,1 of
\cite{2016ApJ...821...44F}. We repeat the
work of \cite{2018A&A...609A..26G} and
\cite{2018A&A...609A..27S} (see more details
in their papers) for $K_{s} \approx18$ stars,
but including new data from the GALACTICNUCLEUS
survey for distances larger than $1.5$ pc to
compute the {\it Nuker} fit. Moreover, we use
the model for the nonNSC emission computed
in the present work to subtract the GB and
NSD contributions from our sample. We use the
NSD Model 2 from Table\,\ref{Tab:nsd_mir_img}.
The resulting best-fit parameters are given
in Table\,\ref{tab:nuker_model}, where we
explore different sources of systematics
error (see Appendix\,\ref{sec:syst_nuker}).
Figure\,\ref{fig:nuker} shows the {\it Nuker}
fit ID\,10 in Table\ref{tab:nuker_model}. The black point represents the corrected surface density data for stars in the magnitude interval $17.5\leq K_{s}\leq 18.5$ from \cite{2018A&A...609A..26G}, complemented at large radii by scaled data from GALACTICNUCLEUS data. In order to obtain the value for the inner power-law $\gamma$ that describes the cusp, we use the mean and standard deviation of the best-fit values of Table\,\ref{tab:nuker_model} and we add the statistical systematic errors quadratically. We obtain a value of $\gamma=1.38\pm0.06_{sys}\pm0.01_{stat}$. As we can see, the mean parameters agree within the uncertainties with the value of $\gamma=1.43\pm0.02\pm0.1_{sys}$ determined in \cite{2018A&A...609A..26G}.

\begin{figure}[!htb]
\includegraphics[width=\columnwidth]{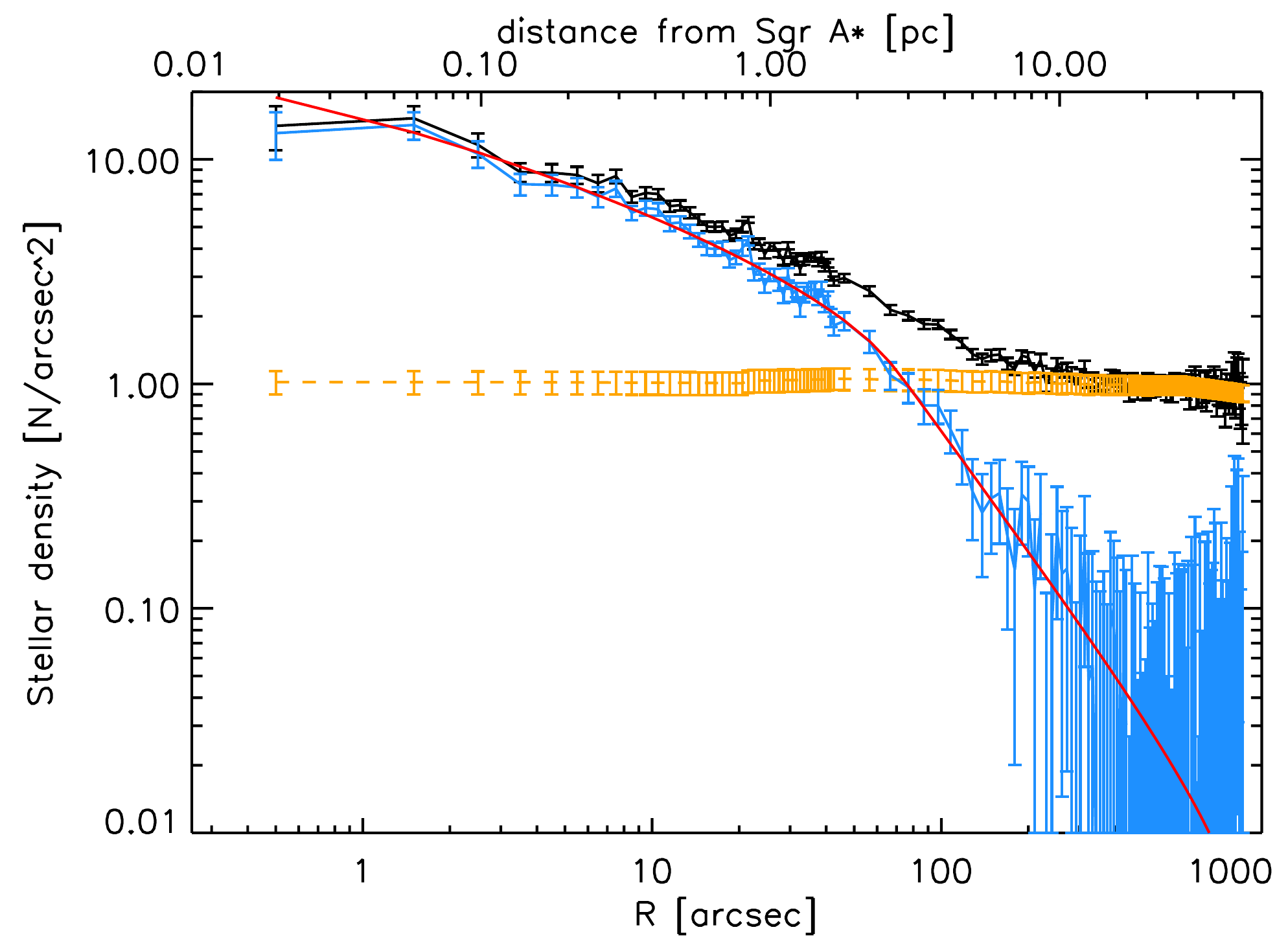}
\caption{\label{fig:nuker} Black: Combined, corrected surface density data for stars in the magnitude interval $17.5\leq K_{s}\leq 18.5$ from \cite{2018A&A...609A..26G}, plus scaled data from the GALACTICNUCLEUS survey at large radii. The blue data points are the data after subtraction of the Model 2 in Table\,\ref{Tab:nsd_mir_img} for the emission from the nuclear stellar disk and the Galactic bulge (the dashed orange line). The red line is a {\it Nuker} model fit (ID\,10 in Table\,\ref{tab:nuker_model}). }
\end{figure}

\subsection{Clues to the Milky Way nuclear star cluster formation}

Nuclear stellar clusters may grow by in situ star formation or by the accretion of other stellar clusters formed sufficiently close to them so that they can be accreted within a Hubble time. Since the relaxation time in NSCs can be as long as a Hubble time \citep[e.g., in the case of the MW, according to][]{2017ARA&A..55...17A}, this implies that accretion or star formation events may be observable in the form of structural or dynamical peculiarities. Such peculiarities have been observed: a starburst happened a few million years ago within $0.5$ pc of the central black hole of the Milky Way \citep[see ][and references therein]{Genzel:2010fk} and also some extragalactic NSCs show centrally concentrated young populations \citep{Georgiev:2014ve,2015AJ....149..170C}. However, younger populations can also be found at greater radii in NSCs \citep{2015AJ....149..170C}. Our observation of the flattened  distribution of bright giants and the observation of a potential rotating substructure perpendicular to the Galactic plane at a projected radius of about 0.8 pc from the center by \cite{Feldmeier:2014kx} indicate that dynamically unrelaxed populations may also be present at larger radii in the MWNSC. While we cannot accurately constrain the formation of the MWNSC from the existing incomplete observational data, those substructures at least provide an indication that it can grow through in situ formation and accretion of nearby clusters. To look into what mechanism plays a more important role in the formation of the MWNSC, we need higher angular resolution NIR imaging and spectroscopy in order to disentangle the different stellar populations inside the NSC.

\section{Conclusions \label{sec:conclusions}}
In order to constrain the overall properties of the NSC of the  Milky Way and overcome some potential shortcomings of previous studies, we analyzed two different datasets and methods. Using star count analyses of high-angular-resolution images from NACO and HAWK-I we created extinction-corrected stellar density maps for old stars. We also improved the analysis of the IRAC/Spitzer MIR images, which are less affected by extinction than those from NACO and HAWK-1. We model the Galactic bulge and the nuclear stellar disk that surround the MWNSC. We subtracted both foreground and background contributions from the MWNSC and fit S\'ersic models to the images. We obtained a value of the effective radius $R_{e}=(5.1 \pm1.0)$ pc. Moreover, we found that the MWNSC is flattened, with its major axis almost parallel to the Galactic plane. The axis ratio is $q=0.71 \pm0.10$, the ellipticity is $\epsilon=1-q=0.29 \pm0.10$, and the S\'ersic index is $n=2.2 \pm0.7$. The structure of the MWNSC is similar to that of other extragalactic NSCs found in spiral galaxies.

We analyzed the symmetry of the MWNSC, firstly assuming that it is aligned with respect to the Galactic plane, in agreement with previous studies, and secondly leaving the angle between the major axis of the MWNSC and the GP free to vary in the fits. We found that the major axis of the MWNSC may be tilted out of the GP by as much as -10 degrees, which is novel confirmation of the value of the angle in the kinematic position obtained by \citet{Feldmeier:2014kx} and of the observations of NSCs from nearby galaxies.

Thanks to our high-angular-resolution images, we were able to study the structure of the MWNSC in different ranges of magnitude. We found some differences between the distributions of brighter giants and RC stars, the former being significantly more flattened. Bright giants are younger than RC stars, on average, according to stellar evolutionary models, and are expected to have bluer colors than the older ones. Furthermore, our findings are in agreement with the differences found between the effective radius of NSCs observed in nearby
galaxies at different wavelengths, which may suggest that the populations of different ages are not completely mixed. The observation of the flattened distribution of bright giants may be support for the in situ star formation scenario for the MWNSC.

Finally, we studied the stellar cusp around Sgr\,A* using GALACTICNUCLEUS data for distances greater than $1.5$ pc and assuming our new GB and NSD models in the {\it Nuker} fit. This cusp is
well developed inside the influence radius of Sgr\,A* and can be described by a single three-dimensional power law with an exponent $\gamma=1.38\pm0.06_{sys}\pm0.01_{stat}$, consistent with previous work.

\begin{acknowledgements}
  The research leading to these results has received funding from the European Research Council under the European Union's Seventh Framework Programme (FP7/2007-2013) / ERC grant agreement n$^{\circ}$[614922]. This work is based on observations made with ESO Telescopes at the La Silla Paranal Observatory under programmes IDs 195.B-0283 and 091.B-0418. We thank the staff of ESO for their great efforts and helpfulness. N.N. acknowledges support by the Sonderforschungsbereich SFB 881 “The Milky Way System” of the German Research Foundation (DFG). F.N.-L. acknowledges financial support from a MECD pre-doctoral contract, code FPU14/01700. We acknowledge support by the State Agency for Research of the Spanish MCIU through the “Center of Excellence Severo Ochoa” award for the Instituto de Astrof\'isica de Andaluc\'ia (CSIC) (SEV-2017-0709).
\end{acknowledgements}

%
%
\bibliography{/Users/lgc/BibDesk/BibGC}

\begin{thebibliography}{74}
\expandafter\ifx\csname natexlab\endcsname\relax\def\natexlab#1{#1}\fi

\bibitem[{{Alexander}(2017)}]{2017ARA&A..55...17A}
{Alexander}, T. 2017, \araa, 55, 17

\bibitem[{{Antonini}(2013)}]{Antonini:2013ys}
{Antonini}, F. 2013, \apj, 763, 62

\bibitem[{{Antonini}(2014)}]{Antonini:2014aa}
{Antonini}, F. 2014, \apj, 794, 106

\bibitem[{{Arca-Sedda} \& {Capuzzo-Dolcetta}(2014)}]{Arca-Sedda:2014aa}
{Arca-Sedda}, M. \& {Capuzzo-Dolcetta}, R. 2014, \apj, 785, 51

\bibitem[{{Athanassoula} {et~al.}(1990){Athanassoula}, {Morin}, {Wozniak},
  {Puy}, {Pierce}, {Lombard}, \& {Bosma}}]{1990MNRAS.245..130A}
{Athanassoula}, E., {Morin}, S., {Wozniak}, H., {et~al.} 1990, \mnras, 245, 130

\bibitem[{{Bahcall} \& {Wolf}(1976)}]{Bahcall:1976vn}
{Bahcall}, J.~N. \& {Wolf}, R.~A. 1976, \apj, 209, 214

\bibitem[{{Bartko} {et~al.}(2010){Bartko}, {Martins}, {Trippe}, {Fritz},
  {Genzel}, {Ott}, {Eisenhauer}, {Gillessen}, {Paumard}, {Alexander},
  {Dodds-Eden}, {Gerhard}, {Levin}, {Mascetti}, {Nayakshin}, {Perets},
  {Perrin}, {Pfuhl}, {Reid}, {Rouan}, {Zilka}, \& {Sternberg}}]{Bartko:2010fk}
{Bartko}, H., {Martins}, F., {Trippe}, S., {et~al.} 2010, \apj, 708, 834

\bibitem[{{Baumgardt} {et~al.}(2018){Baumgardt}, {Amaro-Seoane}, \&
  {Sch{\"o}del}}]{2018A&A...609A..28B}
{Baumgardt}, H., {Amaro-Seoane}, P., \& {Sch{\"o}del}, R. 2018, \aap, 609, A28

\bibitem[{{Becklin} \& {Neugebauer}(1968)}]{Becklin:1968nx}
{Becklin}, E.~E. \& {Neugebauer}, G. 1968, \apj, 151, 145

\bibitem[{{Blum} {et~al.}(1996){Blum}, {Sellgren}, \& {Depoy}}]{Blum:1996mz}
{Blum}, R.~D., {Sellgren}, K., \& {Depoy}, D.~L. 1996, \apj, 470, 864

\bibitem[{{Boehle} {et~al.}(2016){Boehle}, {Ghez}, {Sch{\"o}del}, {Meyer},
  {Yelda}, {Albers}, {Martinez}, {Becklin}, {Do}, {Lu}, {Matthews}, {Morris},
  {Sitarski}, \& {Witzel}}]{Boehle:2016zr}
{Boehle}, A., {Ghez}, A.~M., {Sch{\"o}del}, R., {et~al.} 2016, \apj, 830, 17

\bibitem[{{B{\"o}ker} {et~al.}(2002){B{\"o}ker}, {Laine}, {van der Marel},
  {Sarzi}, {Rix}, {Ho}, \& {Shields}}]{Boker:2002kx}
{B{\"o}ker}, T., {Laine}, S., {van der Marel}, R.~P., {et~al.} 2002, \aj, 123,
  1389

\bibitem[{{B{\"o}ker} {et~al.}(2004){B{\"o}ker}, {Sarzi}, {McLaughlin}, {van
  der Marel}, {Rix}, {Ho}, \& {Shields}}]{Boker:2004oq}
{B{\"o}ker}, T., {Sarzi}, M., {McLaughlin}, D.~E., {et~al.} 2004, \aj, 127, 105

\bibitem[{{Brown} {et~al.}(2018){Brown}, {Gnedin}, \& {Li}}]{Brown:2018aa}
{Brown}, G., {Gnedin}, O.~Y., \& {Li}, H. 2018, \apj, 864, 94

\bibitem[{{Buchholz} {et~al.}(2009){Buchholz}, {Sch{\"o}del}, \&
  {Eckart}}]{Buchholz:2009fk}
{Buchholz}, R.~M., {Sch{\"o}del}, R., \& {Eckart}, A. 2009, \aap, 499, 483

\bibitem[{{Capaccioli}(1987)}]{Capaccioli:1987vn}
{Capaccioli}, M. 1987, in IAU Symposium, Vol. 127, Structure and Dynamics of
  Elliptical Galaxies, ed. P.~T. {de Zeeuw} \& S.~D. {Tremaine}, 47--60

\bibitem[{{Carson} {et~al.}(2015){Carson}, {Barth}, {Seth}, {den Brok},
  {Cappellari}, {Greene}, {Ho}, \& {Neumayer}}]{2015AJ....149..170C}
{Carson}, D.~J., {Barth}, A.~J., {Seth}, A.~C., {et~al.} 2015, \aj, 149, 170

\bibitem[{{Chatzopoulos} {et~al.}(2015){Chatzopoulos}, {Fritz}, {Gerhard},
  {Gillessen}, {Wegg}, {Genzel}, \& {Pfuhl}}]{Chatzopoulos:2015yu}
{Chatzopoulos}, S., {Fritz}, T.~K., {Gerhard}, O., {et~al.} 2015, \mnras, 447,
  948

\bibitem[{{C{\^o}t{\'e}} {et~al.}(2006){C{\^o}t{\'e}}, {Piatek}, {Ferrarese},
  {Jord{\'a}n}, {Merritt}, {Peng}, {Ha{\c s}egan}, {Blakeslee}, {Mei}, {West},
  {Milosavljevi{\'c}}, \& {Tonry}}]{Cote:2006eu}
{C{\^o}t{\'e}}, P., {Piatek}, S., {Ferrarese}, L., {et~al.} 2006, \apjs, 165,
  57

\bibitem[{{Do} {et~al.}(2009){Do}, {Ghez}, {Morris}, {Lu}, {Matthews}, {Yelda},
  \& {Larkin}}]{Do:2009tg}
{Do}, T., {Ghez}, A.~M., {Morris}, M.~R., {et~al.} 2009, \apj, 703, 1323

\bibitem[{{Do} {et~al.}(2015){Do}, {Kerzendorf}, {Winsor}, {St{\o}stad},
  {Morris}, {Lu}, \& {Ghez}}]{Do:2015ve}
{Do}, T., {Kerzendorf}, W., {Winsor}, N., {et~al.} 2015, \apj, 809, 143

\bibitem[{{Do} {et~al.}(2013){Do}, {Lu}, {Ghez}, {Morris}, {Yelda}, {Martinez},
  {Wright}, \& {Matthews}}]{Do:2013fk}
{Do}, T., {Lu}, J.~R., {Ghez}, A.~M., {et~al.} 2013, \apj, 764, 154

\bibitem[{{Dong} {et~al.}(2011){Dong}, {Wang}, {Cotera}, {Stolovy}, {Morris},
  {Mauerhan}, {Mills}, {Schneider}, {Calzetti}, \& {Lang}}]{Dong:2011ff}
{Dong}, H., {Wang}, Q.~D., {Cotera}, A., {et~al.} 2011, \mnras, 417, 114

\bibitem[{{Dwek} {et~al.}(1995){Dwek}, {Arendt}, {Hauser}, {Kelsall}, {Lisse},
  {Moseley}, {Silverberg}, {Sodroski}, \& {Weiland}}]{Dwek:1995uq}
{Dwek}, E., {Arendt}, R.~G., {Hauser}, M.~G., {et~al.} 1995, \apj, 445, 716

\bibitem[{{Feldmeier} {et~al.}(2014){Feldmeier}, {Neumayer}, {Seth},
  {Sch{\"o}del}, {L{\"u}tzgendorf}, {de Zeeuw}, {Kissler-Patig}, {Nishiyama},
  \& {Walcher}}]{Feldmeier:2014kx}
{Feldmeier}, A., {Neumayer}, N., {Seth}, A., {et~al.} 2014, \aap, 570, A2

\bibitem[{{Feldmeier-Krause} {et~al.}(2015){Feldmeier-Krause}, {Neumayer},
  {Sch{\"o}del}, {Seth}, {Hilker}, {de Zeeuw}, {Kuntschner}, {Walcher},
  {L{\"u}tzgendorf}, \& {Kissler-Patig}}]{Feldmeier-Krause:2015}
{Feldmeier-Krause}, A., {Neumayer}, N., {Sch{\"o}del}, R., {et~al.} 2015, \aap,
  584, A2

\bibitem[{{Feldmeier-Krause} {et~al.}(2017){Feldmeier-Krause}, {Zhu},
  {Neumayer}, {van de Ven}, {de Zeeuw}, \&
  {Sch{\"o}del}}]{Feldmeier-Krause:2017rt}
{Feldmeier-Krause}, A., {Zhu}, L., {Neumayer}, N., {et~al.} 2017, \mnras, 466,
  4040

\bibitem[{{Freitag} {et~al.}(2006){Freitag}, {Amaro-Seoane}, \&
  {Kalogera}}]{Freitag:2006fk}
{Freitag}, M., {Amaro-Seoane}, P., \& {Kalogera}, V. 2006, \apj, 649, 91

\bibitem[{{Freudenreich}(1998)}]{1998ApJ...492..495F}
{Freudenreich}, H.~T. 1998, \apj, 492, 495

\bibitem[{{Fritz} {et~al.}(2014){Fritz}, {Chatzopoulos}, {Gerhard},
  {Gillessen}, {Genzel}, {Pfuhl}, {Tacchella}, {Eisenhauer}, \&
  {Ott}}]{Fritz:2014vn}
{Fritz}, T.~K., {Chatzopoulos}, S., {Gerhard}, O., {et~al.} 2014, in IAU
  Symposium, Vol. 303, IAU Symposium, ed. L.~O. {Sjouwerman}, C.~C. {Lang}, \&
  J.~{Ott}, 248--251

\bibitem[{{Fritz} {et~al.}(2016){Fritz}, {Chatzopoulos}, {Gerhard},
  {Gillessen}, {Genzel}, {Pfuhl}, {Tacchella}, {Eisenhauer}, \&
  {Ott}}]{2016ApJ...821...44F}
{Fritz}, T.~K., {Chatzopoulos}, S., {Gerhard}, O., {et~al.} 2016, \apj, 821, 44

\bibitem[{{Fritz} {et~al.}(2011){Fritz}, {Gillessen}, {Dodds-Eden}, {Lutz},
  {Genzel}, {Raab}, {Ott}, {Pfuhl}, {Eisenhauer}, \&
  {Yusef-Zadeh}}]{Fritz:2011fk}
{Fritz}, T.~K., {Gillessen}, S., {Dodds-Eden}, K., {et~al.} 2011, \apj, 737, 73

\bibitem[{{Gallego-Cano} {et~al.}(2018){Gallego-Cano}, {Sch{\"o}del}, {Dong},
  {Nogueras-Lara}, {Gallego-Calvente}, {Amaro-Seoane}, \&
  {Baumgardt}}]{2018A&A...609A..26G}
{Gallego-Cano}, E., {Sch{\"o}del}, R., {Dong}, H., {et~al.} 2018, \aap, 609,
  A26

\bibitem[{{Genzel} {et~al.}(2010){Genzel}, {Eisenhauer}, \&
  {Gillessen}}]{Genzel:2010fk}
{Genzel}, R., {Eisenhauer}, F., \& {Gillessen}, S. 2010, Reviews of Modern
  Physics, 82, 3121

\bibitem[{{Georgiev} \& {B{\"o}ker}(2014)}]{Georgiev:2014ve}
{Georgiev}, I.~Y. \& {B{\"o}ker}, T. 2014, \mnras, 441, 3570

\bibitem[{{Ghez} {et~al.}(2008){Ghez}, {Salim}, {Weinberg}, {Lu}, {Do}, {Dunn},
  {Matthews}, {Morris}, {Yelda}, {Becklin}, {Kremenek}, {Milosavljevic}, \&
  {Naiman}}]{Ghez:2008fk}
{Ghez}, A.~M., {Salim}, S., {Weinberg}, N.~N., {et~al.} 2008, \apj, 689, 1044

\bibitem[{{Gillessen} {et~al.}(2009){Gillessen}, {Eisenhauer}, {Trippe},
  {Alexander}, {Genzel}, {Martins}, \& {Ott}}]{Gillessen:2009qe}
{Gillessen}, S., {Eisenhauer}, F., {Trippe}, S., {et~al.} 2009, \apj, 692, 1075

\bibitem[{{Gillessen} {et~al.}(2017){Gillessen}, {Plewa}, {Eisenhauer}, {Sari},
  {Waisberg}, {Habibi}, {Pfuhl}, {George}, {Dexter}, {von Fellenberg}, {Ott},
  \& {Genzel}}]{2017ApJ...837...30G}
{Gillessen}, S., {Plewa}, P.~M., {Eisenhauer}, F., {et~al.} 2017, \apj, 837, 30

\bibitem[{{Graham}(2001)}]{Graham:2001ys}
{Graham}, A.~W. 2001, \aj, 121, 820

\bibitem[{{Graham} \& {Spitler}(2009)}]{Graham:2009lh}
{Graham}, A.~W. \& {Spitler}, L.~R. 2009, \mnras, 397, 1003

\bibitem[{{Gravity Collaboration} {et~al.}(2018){Gravity Collaboration},
  {Abuter}, {Amorim}, {Anugu}, {Baub{\"o}ck}, {Benisty}, {Berger}, {Blind},
  {Bonnet}, {Brandner}, {Buron}, {Collin}, {Chapron}, {Cl{\'e}net}, {Coud{\'e}
  Du Foresto}, {de Zeeuw}, {Deen}, {Delplancke-Str{\"o}bele}, {Dembet},
  {Dexter}, {Duvert}, {Eckart}, {Eisenhauer}, {Finger}, {F{\"o}rster
  Schreiber}, {F{\'e}dou}, {Garcia}, {Garcia Lopez}, {Gao}, {Gendron},
  {Genzel}, {Gillessen}, {Gordo}, {Habibi}, {Haubois}, {Haug}, {Hau{\ss}mann},
  {Henning}, {Hippler}, {Horrobin}, {Hubert}, {Hubin}, {Jimenez Rosales},
  {Jochum}, {Jocou}, {Kaufer}, {Kellner}, {Kendrew}, {Kervella}, {Kok},
  {Kulas}, {Lacour}, {Lapeyr{\`e}re}, {Lazareff}, {Le Bouquin}, {L{\'e}na},
  {Lippa}, {Lenzen}, {M{\'e}rand}, {M{\"u}ler}, {Neumann}, {Ott}, {Palanca},
  {Paumard}, {Pasquini}, {Perraut}, {Perrin}, {Pfuhl}, {Plewa}, {Rabien},
  {Ram{\'{\i}}rez}, {Ramos}, {Rau}, {Rodr{\'{\i}}guez-Coira}, {Rohloff},
  {Rousset}, {Sanchez-Bermudez}, {Scheithauer}, {Sch{\"o}ller}, {Schuler},
  {Spyromilio}, {Straub}, {Straubmeier}, {Sturm}, {Tacconi}, {Tristram},
  {Vincent}, {von Fellenberg}, {Wank}, {Waisberg}, {Widmann}, {Wieprecht},
  {Wiest}, {Wiezorrek}, {Woillez}, {Yazici}, {Ziegler}, \&
  {Zins}}]{2018A&A...615L..15G}
{Gravity Collaboration}, {Abuter}, R., {Amorim}, A., {et~al.} 2018, \aap, 615,
  L15

\bibitem[{{Hopman} \& {Alexander}(2006)}]{Hopman:2006eu}
{Hopman}, C. \& {Alexander}, T. 2006, \apjl, 645, L133

\bibitem[{{Knuth}(2006)}]{Knuth:2006pK}
{Knuth}, K.~H. 2006, ArXiv Physics e-prints [\eprint{physics/0605197}]

\bibitem[{{Lauer} {et~al.}(1995){Lauer}, {Ajhar}, {Byun}, {Dressler}, {Faber},
  {Grillmair}, {Kormendy}, {Richstone}, \& {Tremaine}}]{Lauer:1995fk}
{Lauer}, T.~R., {Ajhar}, E.~A., {Byun}, Y.-I., {et~al.} 1995, \aj, 110, 2622

\bibitem[{{Launhardt} {et~al.}(2002){Launhardt}, {Zylka}, \&
  {Mezger}}]{Launhardt:2002nx}
{Launhardt}, R., {Zylka}, R., \& {Mezger}, P.~G. 2002, \aap, 384, 112

\bibitem[{{Lightman} \& {Shapiro}(1977)}]{Lightman:1977ly}
{Lightman}, A.~P. \& {Shapiro}, S.~L. 1977, \apj, 211, 244

\bibitem[{{Lu} {et~al.}(2013){Lu}, {Do}, {Ghez}, {Morris}, {Yelda}, \&
  {Matthews}}]{Lu:2013fk}
{Lu}, J.~R., {Do}, T., {Ghez}, A.~M., {et~al.} 2013, \apj, 764, 155

\bibitem[{{McLaughlin} {et~al.}(2006){McLaughlin}, {Anderson}, {Meylan},
  {Gebhardt}, {Pryor}, {Minniti}, \& {Phinney}}]{McLaughlin:2006fk}
{McLaughlin}, D.~E., {Anderson}, J., {Meylan}, G., {et~al.} 2006, \apjs, 166,
  249

\bibitem[{{Meyer} {et~al.}(2012){Meyer}, {Ghez}, {Sch{\"o}del}, {Yelda},
  {Boehle}, {Lu}, {Do}, {Morris}, {Becklin}, \& {Matthews}}]{Meyer:2012fk}
{Meyer}, L., {Ghez}, A.~M., {Sch{\"o}del}, R., {et~al.} 2012, Science, 338, 84

\bibitem[{{Milosavljevi{\'c}} \& {Merritt}(2001)}]{Milosavljevic:2001ly}
{Milosavljevi{\'c}}, M. \& {Merritt}, D. 2001, \apj, 563, 34

\bibitem[{{Molinari} {et~al.}(2011){Molinari}, {Bally}, {Noriega-Crespo},
  {Compi{\`e}gne}, {Bernard}, {Paradis}, {Martin}, {Testi}, {Barlow}, {Moore},
  {Plume}, {Swinyard}, {Zavagno}, {Calzoletti}, {Di Giorgio}, {Elia},
  {Faustini}, {Natoli}, {Pestalozzi}, {Pezzuto}, {Piacentini}, {Polenta},
  {Polychroni}, {Schisano}, {Traficante}, {Veneziani}, {Battersby}, {Burton},
  {Carey}, {Fukui}, {Li}, {Lord}, {Morgan}, {Motte}, {Schuller},
  {Stringfellow}, {Tan}, {Thompson}, {Ward-Thompson}, {White}, \&
  {Umana}}]{Molinari:2011fk}
{Molinari}, S., {Bally}, J., {Noriega-Crespo}, A., {et~al.} 2011, \apjl, 735,
  L33

\bibitem[{{Neumayer} \& {Walcher}(2012)}]{Neumayer:2012fk}
{Neumayer}, N. \& {Walcher}, C.~J. 2012, Advances in Astronomy, 2012
  [\eprint[arXiv]{1201.4950}]

\bibitem[{{Nishiyama} {et~al.}(2008){Nishiyama}, {Nagata}, {Tamura}, {Kandori},
  {Hatano}, {Sato}, \& {Sugitani}}]{Nishiyama:2008qa}
{Nishiyama}, S., {Nagata}, T., {Tamura}, M., {et~al.} 2008, \apj, 680, 1174

\bibitem[{{Nishiyama} {et~al.}(2009){Nishiyama}, {Tamura}, {Hatano}, {Kato},
  {Tanab{\'e}}, {Sugitani}, \& {Nagata}}]{Nishiyama:2009oj}
{Nishiyama}, S., {Tamura}, M., {Hatano}, H., {et~al.} 2009, \apj, 696, 1407

\bibitem[{{Nishiyama} {et~al.}(2013){Nishiyama}, {Yasui}, {Nagata},
  {Yoshikawa}, {Uchiyama}, {Sch{\"o}del}, {Hatano}, {Sato}, {Sugitani},
  {Suenaga}, {Kwon}, \& {Tamura}}]{Nishiyama:2013uq}
{Nishiyama}, S., {Yasui}, K., {Nagata}, T., {et~al.} 2013, \apjl, 769, L28

\bibitem[{{Nogueras-Lara} {et~al.}(2018{\natexlab{a}}){Nogueras-Lara},
  {Gallego-Calvente}, {Dong}, {Gallego-Cano}, {Girard}, {Hilker}, {de Zeeuw},
  {Feldmeier-Krause}, {Nishiyama}, {Najarro}, {Neumayer}, \&
  {Sch{\"o}del}}]{2018A&A...610A..83N}
{Nogueras-Lara}, F., {Gallego-Calvente}, A.~T., {Dong}, H., {et~al.}
  2018{\natexlab{a}}, \aap, 610, A83

\bibitem[{{Nogueras-Lara} {et~al.}(2018{\natexlab{b}}){Nogueras-Lara},
  {Sch{\"o}del}, {Dong}, {Najarro}, {Gallego-Calvente}, {Hilker},
  {Gallego-Cano}, {Nishiyama}, {Neumayer}, {Feldmeier-Krause}, {Girard},
  {Cassisi}, \& {Pietrinferni}}]{2018A&A...620A..83N}
{Nogueras-Lara}, F., {Sch{\"o}del}, R., {Dong}, H., {et~al.}
  2018{\natexlab{b}}, \aap, 620, A83

\bibitem[{{Nogueras-Lara} {et~al.}(2019{\natexlab{a}}){Nogueras-Lara},
  {Sch{\"o}del}, {Gallego-Calvente}, {Dong}, {Gallego-Cano}, {Shahzamanian},
  {Girard}, {Nishiyama}, {Najarro}, \& {Neumayer}}]{2019A&A...631A..20N}
{Nogueras-Lara}, F., {Sch{\"o}del}, R., {Gallego-Calvente}, A.~T., {et~al.}
  2019{\natexlab{a}}, \aap, 631, A20

\bibitem[{{Nogueras-Lara} {et~al.}(2019{\natexlab{b}}){Nogueras-Lara},
  {Sch{\"o}del}, {Gallego-Calvente}, {Gallego-Cano}, {Shahzamanian}, {Dong},
  {Neumayer}, {Hilker}, {Najarro}, {Nishiyama}, {Feldmeier-Krause}, {Girard},
  \& {Cassisi}}]{2019arXiv191006968N}
{Nogueras-Lara}, F., {Sch{\"o}del}, R., {Gallego-Calvente}, A.~T., {et~al.}
  2019{\natexlab{b}}, arXiv e-prints, arXiv:1910.06968

\bibitem[{{Nogueras-Lara} {et~al.}(2019{\natexlab{c}}){Nogueras-Lara},
  {Sch{\"o}del}, {Najarro}, {Gallego-Calvente}, {Gallego-Cano}, {Shahzamanian},
  \& {Neumayer}}]{2019A&A...630L...3N}
{Nogueras-Lara}, F., {Sch{\"o}del}, R., {Najarro}, F., {et~al.}
  2019{\natexlab{c}}, \aap, 630, L3

\bibitem[{{Pfuhl} {et~al.}(2011){Pfuhl}, {Fritz}, {Zilka}, {Maness},
  {Eisenhauer}, {Genzel}, {Gillessen}, {Ott}, {Dodds-Eden}, \&
  {Sternberg}}]{Pfuhl:2011uq}
{Pfuhl}, O., {Fritz}, T.~K., {Zilka}, M., {et~al.} 2011, \apj, 741, 108

\bibitem[{{Philipp} {et~al.}(1999){Philipp}, {Zylka}, {Mezger}, {Duschl},
  {Herbst}, \& {Tuffs}}]{Philipp:1999nx}
{Philipp}, S., {Zylka}, R., {Mezger}, P.~G., {et~al.} 1999, \aap, 348, 768

\bibitem[{{Sch{\"o}del} {et~al.}(2014{\natexlab{a}}){Sch{\"o}del}, {Feldmeier},
  {Kunneriath}, {Stolovy}, {Neumayer}, {Amaro-Seoane}, \&
  {Nishiyama}}]{Schodel:2014fk}
{Sch{\"o}del}, R., {Feldmeier}, A., {Kunneriath}, D., {et~al.}
  2014{\natexlab{a}}, \aap, 566, A47

\bibitem[{{Sch{\"o}del} {et~al.}(2014{\natexlab{b}}){Sch{\"o}del}, {Feldmeier},
  {Neumayer}, {Meyer}, \& {Yelda}}]{Schodel:2014bn}
{Sch{\"o}del}, R., {Feldmeier}, A., {Neumayer}, N., {Meyer}, L., \& {Yelda}, S.
  2014{\natexlab{b}}, Classical and Quantum Gravity, 31, 244007

\bibitem[{{Sch{\"o}del} {et~al.}(2018){Sch{\"o}del}, {Gallego-Cano}, {Dong},
  {Nogueras-Lara}, {Gallego-Calvente}, {Amaro-Seoane}, \&
  {Baumgardt}}]{2018A&A...609A..27S}
{Sch{\"o}del}, R., {Gallego-Cano}, E., {Dong}, H., {et~al.} 2018, \aap, 609,
  A27

\bibitem[{{Sch{\"o}del} {et~al.}(2010){Sch{\"o}del}, {Najarro}, {Muzic}, \&
  {Eckart}}]{Schodel:2010fk}
{Sch{\"o}del}, R., {Najarro}, F., {Muzic}, K., \& {Eckart}, A. 2010, \aap, 511,
  A18+

\bibitem[{{Seth} {et~al.}(2006){Seth}, {Dalcanton}, {Hodge}, \&
  {Debattista}}]{Seth:2006uq}
{Seth}, A.~C., {Dalcanton}, J.~J., {Hodge}, P.~W., \& {Debattista}, V.~P. 2006,
  \aj, 132, 2539

\bibitem[{{Stanek} {et~al.}(1997){Stanek}, {Udalski}, {Szyma{\'N}ski},
  {Ka{\L}u{\.Z}ny}, {Kubiak}, {Mateo}, \&
  {Krzemi{\'N}ski}}]{1997ApJ...477..163S}
{Stanek}, K.~Z., {Udalski}, A., {Szyma{\'N}ski}, M., {et~al.} 1997, \apj, 477,
  163

\bibitem[{{Stolovy} {et~al.}(2006){Stolovy}, {Ramirez}, {Arendt}, {Cotera},
  {Yusef-Zadeh}, {Law}, {Gezari}, {Sellgren}, {Karr}, {Moseley}, \&
  {Smith}}]{Stolovy:2006fk}
{Stolovy}, S., {Ramirez}, S., {Arendt}, R.~G., {et~al.} 2006, Journal of
  Physics Conference Series, 54, 176

\bibitem[{{St{\o}stad} {et~al.}(2015){St{\o}stad}, {Do}, {Murray}, {Lu},
  {Yelda}, \& {Ghez}}]{Stostad:2015qf}
{St{\o}stad}, M., {Do}, T., {Murray}, N., {et~al.} 2015, \apj, 808, 106

\bibitem[{{Tremaine} {et~al.}(1975){Tremaine}, {Ostriker}, \&
  {Spitzer}}]{tremaine75}
{Tremaine}, S.~D., {Ostriker}, J.~P., \& {Spitzer}, Jr., L. 1975, \apj, 196,
  407

\bibitem[{{Walcher} {et~al.}(2005){Walcher}, {van der Marel}, {McLaughlin},
  {Rix}, {B{\"o}ker}, {H{\"a}ring}, {Ho}, {Sarzi}, \&
  {Shields}}]{Walcher:2005ys}
{Walcher}, C.~J., {van der Marel}, R.~P., {McLaughlin}, D., {et~al.} 2005,
  \apj, 618, 237

\bibitem[{{Witzel} {et~al.}(2012){Witzel}, {Eckart}, {Bremer}, {Zamaninasab},
  {Shahzamanian}, {Valencia-S.}, {Sch{\"o}del}, {Karas}, {Lenzen}, {Marchili},
  {Sabha}, {Garcia-Marin}, {Buchholz}, {Kunneriath}, \&
  {Straubmeier}}]{Witzel:2012fk}
{Witzel}, G., {Eckart}, A., {Bremer}, M., {et~al.} 2012, \apjs, 203, 18

\bibitem[{{Yusef-Zadeh} {et~al.}(2012){Yusef-Zadeh}, {Bushouse}, \&
  {Wardle}}]{2012ApJ...744...24Y}
{Yusef-Zadeh}, F., {Bushouse}, H., \& {Wardle}, M. 2012, \apj, 744, 24

\end{thebibliography}


\begin{appendix}


\section{Systematic errors on the 2D fit of the stellar density maps \label{sec:syst_2d}}
In this Appendix we examine some of the  potential sources of systematic
error in the computation of the main NSC parameters.

\subsection{Extinction and completeness \label{sec:err_ext}}
As we see in the Section\,\ref{sec:crowding}, we do not apply any completeness correction to the data but we mask out the pixels where the density is too low (or high) by comparing with their neighbors. Moreover, we study only the fields where the completeness is greater than 50\% in the range of magnitudes considered. 

In order to study the systematic uncertainty derived by the extinction correction we
analyze the density map without any corrections. All the
parameters that we obtain in the fits are consistent with the final
results in Table\,\ref{tab:sersicfits} at the $3\sigma$ level. We
repeat the procedure to study the effect on the final NSC parameters of
the extinction index uncertainty from \cite{2018A&A...620A..83N,2019A&A...630L...3N}
($\bigtriangleup \alpha = 0.08$). We use $\alpha = 2.38$ and $\alpha =
2.22$, respectively. The standard deviations of the NSC parameters are shown in
Table\,\ref{tab:anexoext}. The final uncertainties are included in
Table\,\ref{tab:sersicfits} and Table\,\ref{tab:rotatesersicfits} by
adding quadratically to the uncertainties computed in following sections. The standard deviations in the stellar
number density for the NSD at its effective radius are smaller than $0.01$ in all cases. We can see that the flattening is the parameter that is least affected by this effect.

\begin{table*}[!htb]
\centering
\caption{Test of potential sources of systematic errors in the NSC parameters obtained in the S\'ersic fits due to the selection of a different extinction index ($\alpha = 2.22$, $\alpha = 2.30$ and $\alpha = 2.38$).}
\label{tab:anexoext}
\begin{tabular}{l l l l l l l}
\hline
\hline
ID  & Mag. range  & $\sigma(\theta)$ & $\sigma(N_{nsc})$ & $\sigma(q)$& $\sigma(n)$ & $\sigma(R_{e})$\\
& ($K_{s, extc}$)& (degree) & (stars/bin$^{2}$) & & & (pc)\\

\hline

1 & $12.5-14$ & - & $0.55$ & $0.01$ & $0.18$ & $0.24$ \\
2 & $9-12$ & - & $0.16$ & $0.02$ & $0.21$ & $0.02$ \\
3 & $9-14$ & - & $0.45$ & $0.01$ & $0.02$ & $0.09$ \\
4 & $12.5-14$ & $1.42$ & $0.43$ & $0.01$ & $0.13$ & $0.10$ \\
5 & $9-12$ & $0.40$ & $0.16$ & $0.02$ & $0.19$ & $0.02$ \\
6 & $9-14$ & $0.37$ & $0.44$ & $0.01$ & $0.01$ & $0.07$ \\

\hline
\end{tabular}
\tablefoot{The first three rows corresponding to the fits considering $\theta=0$ (the MWNSC aligned with respect to GP). The last three rows corresponding to the fits where we leave $\theta$ free.}
\end{table*}

We also repeat the whole procedure to create the stellar density maps by considering a
less steep value for the extinction index. We use $\alpha = 2.11\pm0.06$ from
\cite{Fritz:2011fk}. The best-fit parameters are consistent with the final values in
Table\,\ref{tab:sersicfits} at the $1-2\sigma$ level. Although the value of
$\alpha\sim2.30$ from \cite{2018A&A...610A..83N} that we use has been obtained for
the central field ($\sim7 .95'\times3.43'$), \cite{2018A&A...620A..83N}  studied
other regions further out and obtained a very similar value of $\alpha$  that is within
the uncertainties. Therefore, we can expect that the value of the extinction index
is approximately constant for the entire FOV of our data. We conclude that
extinction and completeness are not a significant source of systematic error.

\subsection{Mask \label{sec:masks}}
\begin{figure*}[!htb]
\includegraphics[width=\textwidth]{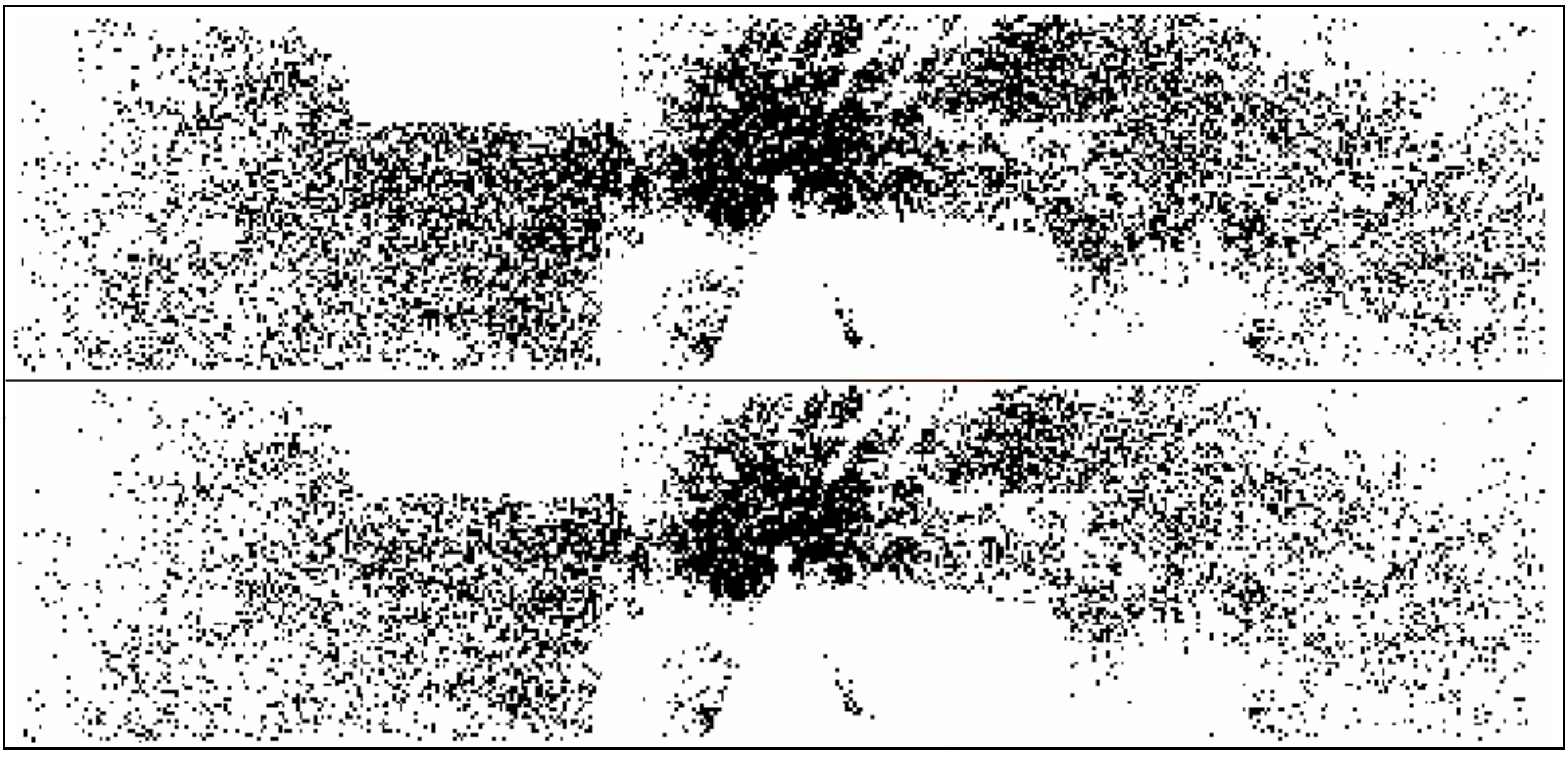}
\caption{\label{Fig:masks} Masks 2 and 3 applied to the extinction-corrected stellar number density maps of the central $86$ pc $\times$ $20.2$ pc of the Galaxy. The masked regions are in white. The differences between both masks are due to the selection of two different values for the density thresholds to mask out regions with low density. Galactic north is up and Galactic east is to the left. The Galactic plane runs horizontally.}
\end{figure*}

In this section, we test the effect of different masks in the fits.
As we explain in section\,\ref{sec:crowding}, we create masks to
take into account the inhomogeneity of the field of HAWK-I data.
Figure \,\ref{Fig:masks} shows two of the masks that we applied.
Masks range from less restrictive (Mask 1) to more restrictive (Mask
4). The differences between the masks are due to the selection of
different values for the density thresholds to mask out regions with
low density (global mask). With the local mask, we also rule out pixels with overly high density. Moreover, we also study the local
mask by rejecting the points whose values are larger than 2-
3$\sigma$, but the differences between the obtained NSC parameters
are negligible.

Table\,\ref{tab:anexosersicfits} shows all the results for masks 1, 2, 3, and 4. The uncertainties are the standard deviations of the best-fit parameters. The selection of the mask is an important source of systematic error that we have already considered in Table\,\ref{tab:sersicfits}.
\begin{table*}[!htb]
\centering
\caption{Test of potential sources of systematic errors in the NSC parameters obtained in the S\'ersic fits due to the selection of the masks. The MWNSC is aligned with respect to GP.}
\label{tab:anexosersicfits}
\begin{tabular}{l l l l l l l l l l l}
\hline
\hline
ID  & Mag. range  & Mask & $N^{b}_{e,nsd}$ & $N^{c}_{e,nsc}$ & $q^{d}$ & $n^{e}$ & $R^{f}_{e}$ \\
& ($K_{s, extc}$)& & (stars/bin$^{2}$) & (stars/bin$^{2}$) & & & (pc) \\
\hline
1 & $12.5-14$ & $3$ & $2.39\pm0.01$ & $3.49\pm0.27$ & $0.83\pm0.01$ & $2.73\pm0.11$ & $5.83\pm0.28$ \\
2 & $12.5-14$ & $4$ & $2.57\pm0.01$ & $3.86\pm0.28$ & $0.79\pm0.01$ & $2.73\pm0.10$ & $5.42\pm0.24$ \\
3 & $12.5-14^{a}_{sym}$ & $1$ & $2.36\pm0.01$ & $4.32\pm0.19$ & $0.89\pm0.01$ & $2.38\pm0.07$ & $5.68\pm0.15$ \\
4 & $12.5-14^{a}_{sym}$ & $2$ & $2.49\pm0.01$ & $4.09\pm0.20$ & $0.87\pm0.01$ & $2.52\pm0.07$ & $5.73\pm0.17$ \\
5 & $9-12$ & $3$ & $0.47\pm0.00$ & $1.25\pm0.13$ & $0.59\pm0.01$ & $2.22\pm0.17$ & $3.92\pm0.23$ \\
6 & $9-12$ & $4$ & $0.49\pm0.00$ & $1.21\pm0.13$ & $0.56\pm0.01$ & $2.25\pm0.17$ & $4.09\pm0.25$ \\
7 & $9-12^{a}_{sym}$ & $1$ & $0.61\pm0.01$ & $1.26\pm0.09$ & $0.62\pm0.01$ & $2.04\pm0.11$ & $4.68\pm0.19$ \\
8 & $9-12^{a}_{sym}$ & $2$ & $0.64\pm0.01$ & $0.91\pm0.09$ & $0.61\pm0.01$ & $2.61\pm0.16$ & $5.61\pm0.35$ \\
9 & $9-14$ & $3$ & $3.23\pm0.01$ & $5.52\pm0.28$ & $0.78\pm0.01$ & $2.43\pm0.07$ & $5.54\pm0.17$ \\
10 & $9-14$ & $4$ & $3.46\pm0.01$ & $5.99\pm0.28$ & $0.75\pm0.01$ & $2.40\pm0.06$ & $5.20\pm0.15$ \\
11 & $9-14^{a}_{sym}$ & $1$ & $2.99\pm0.01$ & $7.52\pm0.16$ & $0.80\pm0.01$ & $1.60\pm0.03$ & $5.40\pm0.07$ \\
12 & $9-14^{a}_{sym}$ & $2$ & $3.19\pm0.01$ & $7.17\pm0.16$ & $0.79\pm0.01$ & $1.63\pm0.03$ & $5.37\pm0.07$ \\
\hline
\end{tabular}
\tablefoot{
\tablefoottext{a}{Fits to the symmetrized images that we create by assuming symmetry of the NSC with respect to the GP and with respect to the Galactic north--south axis through Sgr\,A*. }\\
\tablefoottext{b}{Stellar number density for the NSD at its effective radius. }\\
\tablefoottext{c}{Stellar number density for the NSC at the effective radius $R_{e}$. }\\
\tablefoottext{d}{The flattening $q$ is equal to the minor axis divided by the major axis. }\\
\tablefoottext{e}{$n$ is the S\'ersic index. }\\
\tablefoottext{f}{$R_{e}$ is the effective radius for the NSC.}
}
\end{table*}

We test the systematic errors associated with the selection of the masks for the fits where we leave the tilt angle between the NSC and the GP free to vary. This is an important source of systematic error that we have already considered in Table\,\ref{tab:rotatesersicfits}, which shows all the obtained results. The parameters do not show a systematic behavior if we apply a more restrictive mask.

\begin{table*}[!htb]
\centering
\caption{Test of potential sources of systematic error in the NSC parameters obtained in the S\'ersic fits due to the selection of the masks. The tilt angle $\theta$ between the NSC and the GP is a free parameter in the fits.}
\label{tab:anexosersicrotatefits}
\begin{tabular}{l l l l l l l l l l l l}
\hline
\hline
ID  & Mag. range  & Mask & $\theta^{b}$ & $N_{e,NSD}$ & $N_{e,NSC}$ & $q$ & $n$ & $R_{e}$ \\
& ($K_{s, extc}$)&  & (degree) & (stars/bin$^{2}$) & (stars/bin$^{2}$) & & & (pc) \\
\hline

1 & $12.5-14$ & $3$ & $-8.50\pm1.03$ & $2.39\pm0.01$ & $3.53\pm0.26$ & $0.82\pm0.01$ & $2.70\pm0.10$ & $5.83\pm0.26$ \\
2 & $12.5-14$ & $4$ & $-8.52\pm0.83$ & $2.56\pm0.01$ & $3.87\pm0.28$ & $0.77\pm0.01$ & $2.71\pm0.10$ & $5.45\pm0.24$ \\
3 & $12.5-14^{a}_{sym}$ & $1$ & $-8.51\pm0.00$ & $2.32\pm0.01$ & $3.18\pm0.24$ & $0.87\pm0.01$ & $3.20\pm0.14$ & $6.69\pm0.30$ \\
4 & $12.5-14^{a}_{sym}$ & $2$ & $-8.51\pm0.00$ & $2.45\pm0.01$ & $2.92\pm0.25$ & $0.87\pm0.01$ & $3.43\pm0.16$ & $6.80\pm0.35$ \\
5 & $9-12$ & $3$ & $-5.65\pm0.87$ & $0.47\pm0.01$ & $1.33\pm0.12$ & $0.57\pm0.01$ & $2.06\pm0.15$ & $3.85\pm0.20$ \\
6 & $9-12$ & $4$ & $-5.56\pm0.74$ & $0.49\pm0.01$ & $1.28\pm0.12$ & $0.54\pm0.01$ & $2.12\pm0.15$ & $4.04\pm0.23$ \\
7 & $9-12^{a}_{sym}$ & $1$ & $-5.60\pm0.00$ & $0.60\pm0.01$ & $1.26\pm0.07$ & $0.75\pm0.01$ & $1.67\pm0.08$ & $4.42\pm0.15$ \\
8 & $9-12^{a}_{sym}$ & $2$ & $-5.60\pm0.00$ & $0.63\pm0.01$ & $1.08\pm0.08$ & $0.71\pm0.01$ & $1.87\pm0.10$ & $4.94\pm0.22$ \\
9 & $9-14$ & $3$ & $-7.47\pm0.60$ & $3.23\pm0.01$ & $5.60\pm0.27$ & $0.76\pm0.01$ & $2.39\pm0.07$ & $5.53\pm0.16$ \\
10 & $9-14$ & $4$ & $-6.49\pm0.54$ & $3.45\pm0.01$ & $5.92\pm0.27$ & $0.74\pm0.01$ & $2.40\pm0.06$ & $5.27\pm0.15$ \\
11 & $9-14^{a}_{sym}$ & $1$ & $-6.98\pm0.01$ & $2.96\pm0.00$ & $6.88\pm0.25$ & $0.75\pm0.01$ & $2.23\pm0.07$ & $5.69\pm0.12$ \\
12 & $9-14^{a}_{sym}$ & $2$ & $-6.98\pm0.01$ & $3.18\pm0.00$ & $8.14\pm0.23$ & $0.74\pm0.01$ & $2.00\pm0.05$ & $4.97\pm0.08$ \\

\hline
\end{tabular}
\tablefoot{
\tablefoottext{a}{Fits to the symmetrized images that we create by assuming symmetry of the NSC with respect to a rotated axis with the GP. The value of the tilt angles are the mean of the values obtained in the fits for the images applying two different masks.}\\
\tablefoottext{b}{The tilt angle between the NSC and the GP in degrees. Positive angles are clockwise with respect to the GP.}}
\end{table*}

\subsection{Inner mask}
In this section, we examine several potential sources of systematic error in the NSC parameters due to masking different central regions around Sgr\,A*. In order to explore the systematic errors, we test the effects of masking the central $0.4$ pc, $0.6$ pc, and $0.8$ pc, respectively. Table\,\ref{tab:anexoinnermask} shows the results. The final uncertainties are included in Tables\,\ref{tab:sersicfits} and\,\ref{tab:rotatesersicfits} by adding quadratically to the uncertainties due to the selection of the mask computed in the previous section.

\begin{table*}[!htb]
\centering
\caption{Test of potential sources of systematic errors in the NSC parameters obtained in the S\'ersic fits due to masking the inner $0.4$ pc, $0.6$ pc, and $0.8$ pc, respectively, around Sgr\,A*. We only show the parameters of the magnitude ranges that are significantly affected.}
\label{tab:anexoinnermask}
\begin{tabular}{l l l l l l l l l l}
\hline
\hline
ID  & Mag. range  & $\theta^{b}$ & $\sigma(\theta)$ & $N_{nsc}$ & $\sigma(N_{nsc})$ & $n$ & $\sigma(n)$ & $R_{e}$ & $\sigma(R_{e})$\\
& ($K_{s, extc}$)& (degree) & (degree) & (stars/bin$^{2}$) & (stars/bin$^{2}$) & & & (pc) & (pc)\\

\hline

1 & $12.5-14$ & 0 & - & $3.84,3.86,3.66$ & $0.11$ & $2.75,2.73,2.86$ & $0.07$ & $5.44,5.42,5.57$ & $0.08$ \\
2 & $9-12$ & 0 & - & $1.11,1.21,1.35$ & $0.12$ & $2.53,2.25,1.85$ & $0.34$ & $4.22,4.09,3.92$ & $0.15$ \\
3 & $9-14$ & 0 & - & $5.78,5.99,6.24$ & $0.23$ & $2.47,2.40,2.33$ & $0.07$ & $5.30,5.20,5.06$ & $0.12$\\
4 & $12.5-14$ & $-5.5,-8.5,-10.8$ & $2.6$ & $3.84,3.87,3.96$ & $0.06$ & $2.73,2.71,2.69$ & $0.02$ & $5.47,5.45,5.38$ & $0.05$\\
5 & $9-12$ & $-3.8,-5.6,-6.2$ & $1.2$ & $1.18,1.28,1.48$ & $0.15$ & $2.42,2.12,1.61$ & $0.41$ & $4.14,4.04,3.82$ & $0.16$\\
6 & $9-14$ & $-5.6,-6.5,-7.3$ & $0.9$ & $5.75,5.92,6.33$ & $0.30$ & $2.47,2.40,2.28$ & $0.09$ & $5.34,5.27,5.06$ & $0.14$\\

\hline
\end{tabular}
\tablefoot{The first three rows correspond to the fits considering $\theta=0$ (the MWNSC aligned with respect to GP). The last three rows correspond to the fits where we leave $\theta$ free.}
\end{table*}

\subsection{Spatial binning \label{sec:binning}}

\begin{figure}[!htb]
\includegraphics[width=\columnwidth]{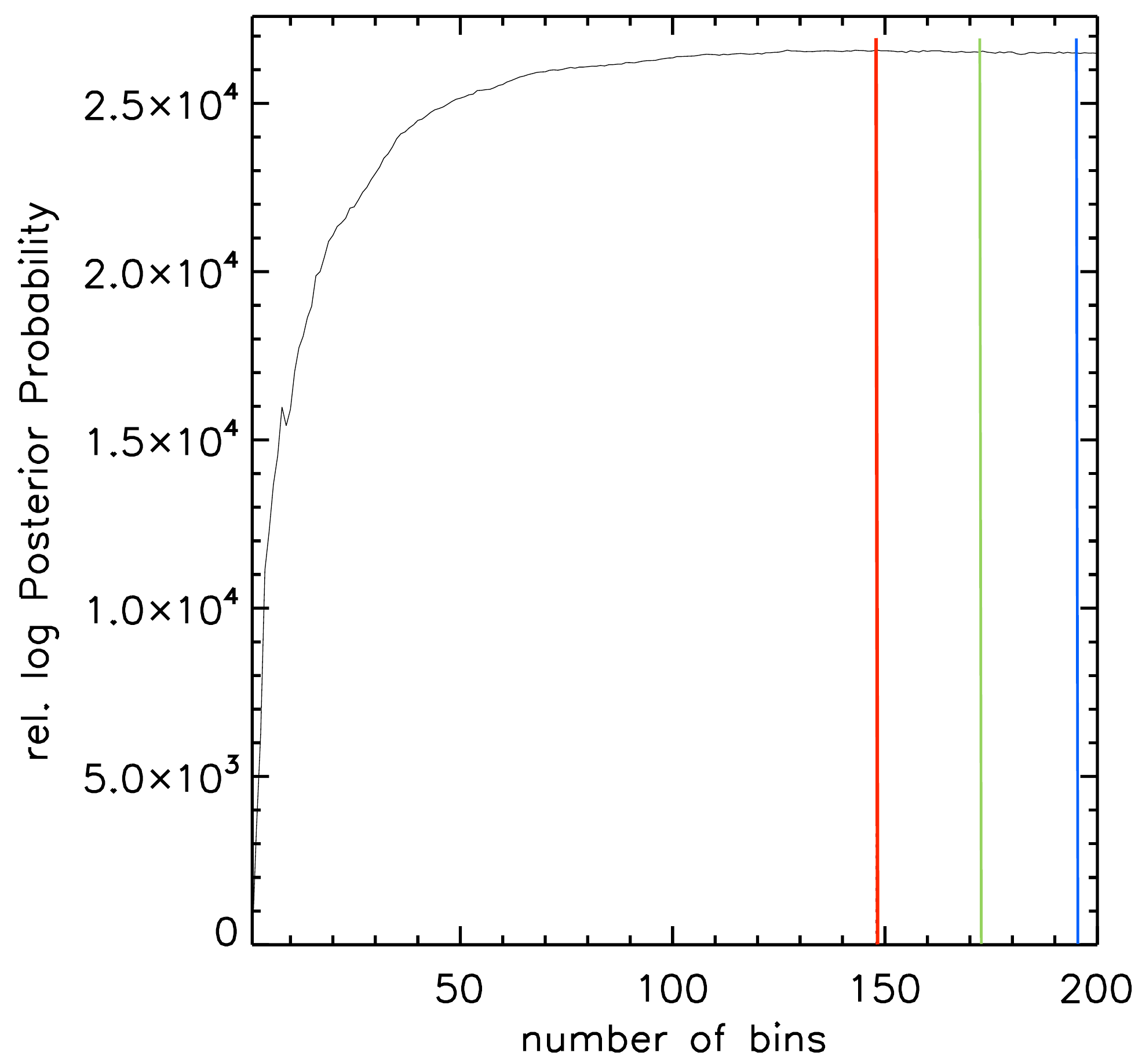}
\caption{\label{Fig:optbin}Optimal spatial binning. We show the RLP
  as a function of the number of bins. The maximum for the
  RLP is reached for 147 bins (red line). The values for 172 bins (green line) and 195 bins (blue line) are obtained for bin sizes of $6''$ and
  $5''$ , respectively. We study all stars brighter than $K_{s}=16.0$}
\end{figure}

We analyzed the results considering different ways of spatially binning the
data. Firstly, we studied the optimal bin size for our density maps following the
studies of \citet{Knuth:2006pK} and
\citet{Witzel:2012fk}. These latter authors applied
Bayesian probability theory to develop a
straightforward algorithm to compute the
posterior probability of bins for a given
data set (see more details in their papers).
We limit our sample to the stars brighter
than $K_{s}=16.0$. The dependence of the relative
logarithmic posterior probability (RLP) on
the bin number is shown in
Fig.\,\ref{Fig:optbin}. The maximum for the
RLP for the
star number is reached for 147 bins, and the
best bin size is $7''$. Secondly, in order to
smooth the overlapping region between data
of different resolution, we test other
smaller values of binning of $5''$ and
$6''$, obtaining the values for the maximum of
the RLP of 195 and 172, respectively. In
figure\,\ref{Fig:optbin}, we can see that the
value of the RLP is very similar for those
values. Regardless, we study our results by
selecting a bin size of $7"\times7"$ and we
obtain all the values for the NSC parameters consistent with the final results in Table\,\ref{tab:sersicfits} at the 2$\sigma$ level. We select a bin size of $5"\times5"$
because it is better suited to our data and allows us to study
the major features at the large scale of the NSC with
enough detail without taking into account the random
fluctuations. We conclude that spatial binning is not
a significant source of systematic error in our analysis.

\subsection{Edges between NACO and HAWK-I}
As we saw in section\,\ref{sec:Methodology}, we compute the mean density at the edges between HAWK-I and NACO data with an outer radius of around $1.5$ pc and inner radius of around $1.3$ pc (see Fig. 3). We explore the sources of systematic error associated with the size of the edges that we choose. We test different values for the inner radius ($1.2$ pc and $1.4$ pc, respectively). The differences between the obtained NSC parameters and the values in Table\,\ref{tab:sersicfits} are negligible.
\subsection{Contamination by stars in the
GD\label{sec:foreground}}
We removed the contribution of NSD and GB in the
NSC fits. We also explored the effect of excluding foreground stars from
the GD based on their color H - K$_{s}$. If we do not
consider stars with $H-K_{s}<1.2$, the best-fit parameters
are consistent with the final values in Table\,\ref{tab:sersicfits} at the $1\sigma$
level.
Therefore, contamination from stars in the GD was found to have no significant impact on the
results. In
addition, some of the excluded stars may be
part of the GB or NSD and we have already taken
them into account; we therefore prefer not to apply this correction.
\subsection{Magnitude cut}
In order to study the RC distribution, we considered stars with K$_{s}$ up to 16
(K$_{s, extc}\sim14.0$), as can be seen in the K$_{s}$-luminosity function from
Fig. \,\ref{Fig:KLF_figure}. However, we analyzed the fits obtained when creating the density maps considering a
magnitude cut of K$_{s}= 15.5$ before the extinction correction. All the values for the
NSC parameters are consistent with the final results in Table\,\ref{tab:sersicfits}
at the $2\sigma$ level.
\subsection{Saturation of stars \label{sec:saturation}}
We select the brightest magnitude in the range of the stars to be not significantly affected by saturation in the images. For HAWK-I data, stars with $K_{s}\sim11.0$ ($K_{s, extc}\sim9.0$) are saturated. We selected the bright limit of $K_{s}=11.0$ ($K_{s, extc}=9.0$) in our analysis, but we also explored the effects of using another bright limit, $K_{s}=10.0$ ($K_{s, extc}=8.0$). The differences in the NSC parameters that we obtain from using these two limits are very small ($\Delta Re <3.1$\%, $\Delta n < 2.5$\%, and $\Delta q <0.7$\%).
\subsection{Possible contamination from young stars}
We aim to study the structure of the relaxed stars in the NSC. We know that a star
formation event created on the order of $10^{4}$\,M$_{\odot}$ of young
stars in the $\sim$  0.5\,pc region  around Sgr\,A* \citep[]{Bartko:2010fk,Lu:2013fk,Feldmeier-Krause:2015}. As we see in section\,\ref{sec:Methodology}, we exclude all spectroscopically identified early-type stars from our sample \citep[using the data of][]{Do:2013fk}. In addition, 90\% of the young stars found in \cite{Feldmeier-Krause:2015} are identified within $0.5$ pc; therefore, further out we do not expect important contamination from young stars in our sample. We apply an inner mask to the $0.6$ pc
region around Sgr\,A*. We conclude that the contamination of young stars in our data is not significant.

\subsection{Initial tilt angle \label{sec:Initial_rotation}}
\begin{figure*}[!htb]
\includegraphics[width=\textwidth]{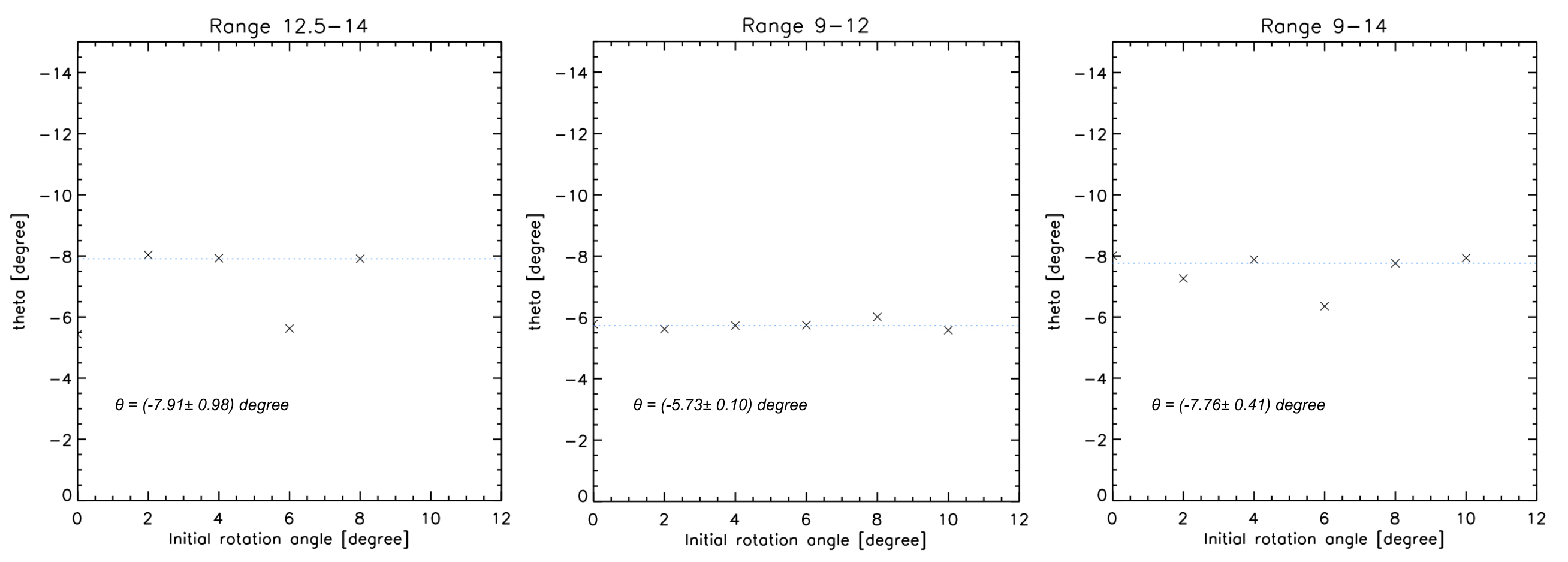}
\caption{\label{Fig:initial_angle} Mean values of the best-fit tilt angles ($\theta$) for the different magnitude ranges obtained from the fits by considering different initial values. The dotted blue line indicates the median $\theta$ and their values are shown in the different panels.}
\end{figure*}

In this section, we focus on the sources of systematic error that affect the possible tilt angle ($\theta$) between the NSC and the GP obtained in the fits. We have studied the effect of the mask in the previous section. In this section, we explore the systematic errors associated with the selection of different initial $\theta$ when performing the fit. As we see in previous sections, in order to find the best-fit parameters, we use the IDL MPCURVEFIT procedure that performs a Levenberg-Marquardt least-squares fit. The algorithm finds the local minimum that minimizes the chi-square, which is not the global minimum in most cases. We compare the average of the residuals obtained by subtracting the symmetrized maps from the images (see Fig\,\ref{Fig:residual}) and we find multiple minima for different tilt angles. For this reason, it is very important to provide an initial guess for $\theta$ close to the final solution in order to obtain the global minimum. Figure\,\ref{Fig:initial_angle} shows the best-fit tilt angles versus different initial values of the angle for the different magnitude ranges. The median values are indicated in the plots. The errors are the average deviation from the median. We reject data more than $3\sigma$ from the median. We conclude that the initial $\theta$ is a significant source of systematic error in our analysis and we include them in our final budget in Table\,\ref{tab:rotatesersicfits}.

\section{Systematic errors of the inner power-law index \label{sec:syst_nuker}}
In this section, we examine several potential sources of systematic error in the computation of the power-law indices for the fits of the 3D density. We only focus on the possible systematic error due to the computation of the star density at large distances  using data from the present work. Table\,\ref{tab:nuker_model} shows the best-fit parameters for the break radius $r_{b}$, for the exponents of the inner, $\gamma$,  and the outer, $\beta$ power-law. We test two ways of binning the data (5 arcsec and 10 arcsec, respectively), use two different masks, and assume two different ranges of magnitudes for the stars that we select (bright and faint stars). Finally, we compute the density along the GP in a strip $\pm50''$ and azimuthally averaged in annuli around Sgr\,A*.

\begin{table*}[!htb]
\centering
\caption{Best-fit model parameters for {\it Nuker} fits.}
\label{tab:nuker_model}
\begin{tabular}{l l l l l l l l }
\hline
\hline
ID   & Mag. range  & Mask & Bin size &$r_{b}$ & $\gamma$ & $\beta$ & $\chi^{2}_{reduced}$\\
&  ($K_{s, extc}$) & & (arcsec)& (pc)& & & \\
\hline
1$^{a}$ & $12.5-14$ & $3$ & $10$ & $1.99\pm0.19$ & $1.34\pm0.05$ & $2.64\pm0.10$ & $1.39$\\
2$^{a}$ & $12.5-14$ & $4$ & $10$ & $1.97\pm0.20$ & $1.34\pm0.05$ & $2.70\pm0.12$ & $1.16$\\
3$^{a}$ & $12.5-14$ & $3$ & $5$ & $1.86\pm0.12$ & $1.30\pm0.05$ & $2.75\pm0.08$ & $1.30$\\
4$^{a}$ & $9-12$ & $3$ & $10$ & $2.23\pm0.47$ & $1.45\pm0.05$ & $2.38\pm0.16$ & $0.64$\\
5$^{a}$ & $9-12$ & $4$ & $10$ & $2.04\pm0.45$ & $1.44\pm0.05$ & $2.33\pm0.15$ & $0.51$\\
6$^{a}$ & $9-12$ & $3$ & $5$ & $2.11\pm0.32$ & $1.44\pm0.04$ & $2.51\pm0.14$ & $0.70$\\
7$^{b}$ & $12.5-14$ & $3$ & $10$ & $2.04\pm0.15$ & $1.33\pm0.05$ & $2.81\pm0.08$ & $1.82$\\
8$^{b}$ & $12.5-14$ & $4$ & $10$ & $2.00\pm0.15$ & $1.32\pm0.05$ & $2.86\pm0.10$ & $1.60$\\
9$^{b}$ & $12.5-14$ & $3$ & $5$ & $1.95\pm0.11$ & $1.30\pm0.05$ & $2.93\pm0.07$ & $1.67$\\
10$^{b}$ & $9-12$ & $3$ & $10$ & $2.67\pm0.37$ & $1.44\pm0.04$ & $2.80\pm0.19$ & $1.00$\\
11$^{b}$ & $9-12$ & $4$ & $10$ & $2.63\pm0.39$ & $1.44\pm0.04$ & $2.87\pm0.23$ & $0.85$\\
12$^{b}$ & $9-12$ & $3$ & $5$ & $2.54\pm0.25$ & $1.42\pm0.04$ & $3.01\pm0.18$ & $0.96$\\
\hline
\end{tabular}
\tablefoot{
\tablefoottext{a}{Fit range: $0.04 \leq R\leq20$ pc. Stellar number density for large radii computed along the GP in a strip $\pm50$ arcsec.}\\
\tablefoottext{b}{Fit range: $0.04 \leq R\leq20$ pc. Stellar number density for large radii has been computed azimuthally averaged in annuli around Sgr\,A*.}
}
\end{table*}

\end{appendix}

\end{document}